\documentclass[%
 reprint,
 amsmath,amssymb,
 aps,pra
]{revtex4-2}

\usepackage[T1]{fontenc}
\makeatletter
\@ifundefined{DeclareUnicodeCharacter}{}{%
  \DeclareUnicodeCharacter{03B3}{\ensuremath{\gamma}}%
  \DeclareUnicodeCharacter{1D62}{\ensuremath{_i}}%
  \DeclareUnicodeCharacter{1D40}{\ensuremath{^\top}}%
  \DeclareUnicodeCharacter{2014}{---}%
  \DeclareUnicodeCharacter{2013}{--}%
  \DeclareUnicodeCharacter{2212}{-}%
  \DeclareUnicodeCharacter{00B7}{\ensuremath{\cdot}}%
  \DeclareUnicodeCharacter{2022}{\textbullet}%
}
\makeatother
\usepackage{amsmath,amssymb,amsthm}
\usepackage{mathtools}
\usepackage{bm}
\usepackage{hyperref}
\hypersetup{hidelinks, breaklinks=false}
\usepackage{cleveref}
\usepackage{booktabs}
\usepackage{graphicx}

\usepackage{natbib}
\newtheorem{theorem}{Theorem}[section]

\newtheorem{definition}[theorem]{Definition}

\newcommand{\R}{\mathbb{R}}

\newcommand{\calM}{\mathcal{M}}
\newcommand{\calA}{\mathcal{A}}

\newcommand{\calB}{\mathcal{B}}

\newcommand{\calX}{\mathcal{X}}
\newcommand{\calY}{\mathcal{Y}}
\newcommand{\dGW}{d_{\mathrm{GW}}}

\newcommand{\dvec}{\mathbf{x}}
\begin{document}

\title{Phase spaces of phase spaces: A method of constructing generalizable and topologically informed bases from time series demonstrated by reconstruction of the geometry of the macroeconomy}

\author{Maximilian Topel}
 \affiliation{Department of Applied Mathematics, Northwestern University, 2145 Sheridan Road, Evanston, Illinois 60208, USA}

\date{\today}
\begin{abstract}

The existence of laws from physics to social science is predicated on some small and enumerable number of factors explaining variation in the systems they describe. While the orthodoxy of economics, physics and chemistry relies on a small number of analytically interpretable laws, large data models pursuing universality throw in all available data. Both fail in their extremes through the same flaw, an incorrect basis: too few orthogonal variables to describe system variation, or an explosion of dimensionality around fundamentally low dimensional but unknown dynamics. In this work we propose and validate a framework that extracts an interpretable, mathematically sound, low dimensional and dynamically informed basis directly from arbitrary timeseries. Combining Takens' Delay Embedding Theorem, dimensionality reduction and optimal transport, we produce a phase space of phase spaces (PSoPS) on which all of our interrelated dynamical series exist: barycenters formed from many phase spaces, together with the families of transport plans between individual phase spaces and the barycenter, yield a basis describing the relational geometry of each timeseries and the topology of economic relationships. The macroeconomic PSoPS built from 281,536 series is a single connected object of effective dimension $7.3 \pm 0.2$, into which the canonical laws (Phillips, Okun, Solow) embed as $\approx$ 2-dimensional looped sub-attractors. The coupling between concept barycenters is directed, with the price of money and productivity as net sources and credit and output as sinks, recovering monetary transmission from geometry alone. IAAFT phase-randomized surrogates destroy 36--43\% of the dimension across the three laws and 38\% on the full corpus, so the basis is irreducibly nonlinear. We close with proposed extensions to fields whose dynamics share attractor geometries and to systems that are intrinsically low dimensional yet difficult to navigate in their raw coordinates.
\end{abstract}

\maketitle

\section{Introduction}

Relatively few models describe macroeconomic relations from the Solow Model \cite{solow1956contribution} to Okun's law \cite{okun1962potential}. At the core of these models is that the macroeconomy is intrinsically low dimensional. If the Federal Reserve changes rates, this has knock-on effects in the global sovereign credit market. It impacts interbank lending rates (e.g.\ SOFR) and the price of corporate debt. It propagates down to mortgage rates and personal credit rates and consequently to default rates and consumption. While there is a clear primary mode here (the price of money), there are other conserved modes. The macroeconomy is a dynamic network driven by low dimensional laws with its degrees of freedom coupled via a high degree of interrelation and cooperation across the global economy. Oil prices and cooperate oil bonds are not independent of one another. If a drought ruins the coffee bean harvest in Ethiopia or transportation issues block natural gas export from Qatar these effects must be felt throughout the world from the farms of Colombia to the chilly lofts of British graduate students. Transformer models have been used to ingest these relevant timeseries and scale to divine their relations. These methods inherit basis-quality from their input coordinates. The objective of this paper is to construct the inputs in a mathematically interpretable, compact space of well defined metric.

This problem is very old. The central object of classical mechanics, certainly as put forth by Lagrange and Hamilton, was to write down the equations of motion describing the trajectory of a system as it moves through a phase space spanned by variables describing the degrees of freedom of the system. This program, applied from kinematics to electrostatics and the multiplexing of everything in between, has been so successful that the same procedure that works for solving the trajectory of a pendulum swinging back and forth can be used to describe the folding and unfolding of a protein from its constituent atomic degrees of freedom. The interaction of each and every atom in protein and solvent is integrated with Hamilton's mechanics to form a complete phase space at the cost of a few (or many) hours of GPU time. Of course the entirety of the world is not integrable in such a fashion so it was precisely in this field of molecular mechanics that the molecular dynamics of Brownian motion were described with stochastic calculus, leaving the deterministic world where a basis is completely described for the world of probabilities. Many fields applied the same approach, most notably here Black and Scholes who carried the diffusion equation of physics into options pricing~\cite{black1973pricing}. But the rules of interaction of degrees of freedom describing a system are not necessarily forthcoming even in the odd case where the basis is. 

This brings us to the world of Random Matrix Theory (RMT) where the confluence of observables and their correlation matrices can be used to extract an eigenspectrum with some guarantee of the significance of the largest modes exceeding a given noise limit. In this world, a large volume of data without guarantees of independence of observable still yields an empirically useful if not readily interpretable basis. This approach found great success in finance and other fields but suffers in a few key ways. The cost of operating on such matrices grows like $N^2$, yes but also the basis vectors they produce are neither inherently interpretable nor do they enjoy any guarantees about a relationship to the dynamics that produces them. Covarying but non-causal origins of phenomena (i.e. phylogenetic vs functional origins of conservation in protein dynamics \cite{halabi2009sectors}) aren't readily parsed. An insufficient sampling of data will still give you eigenvectors, they just may only span some small part of the total variation space that produces the phenomena you study. And worst, for the dynamicist, one can construct such correlation matrices free of dynamical observations (i.e structural observations), resulting in a basis that not only does not describe the dynamics of the system you are studying but may lead you to believe you have understood it!

It is on the shoulders of these two great innovations that the field of random dynamical systems is able to provide an alternative, with some of the interpretability of stochastic calculus and the data driven value of random matrix theory. The central object of this work is to provide a data driven algorithm capable of producing a basis that correctly describes the dynamics of the system being studied. This is done through the marriage of embedding theorems that guarantee the relationship between time series observables and driving dynamics and transport theory that describes the relationship between these latent spaces. 

\subsection{An empirical view of the problem of economic model construction}

In this work, we use the general problem of macroeconomic modeling as the model system for constructing these phase spaces of phase spaces (PSoPS) that serve as a basis for complex and coupled dynamical systems lacking a clear, low dimensional description to span those dynamics. We elect to do so because the field offers tremendous public data, a rich literature with well described and recapitulated phenomenologies but is so complex and high dimensional in the pure observable space that it resists both the stochastic differential equation and random matrix theory approaches. That said, understanding the macroeconomy is so desirable a goal that a plethora of methods have been tried in the attempt.

With macroeconomic timeseries abundant, the space of possible regressions grows with each day. The landmark Fama-French model \cite{famafrench1993} describing stock market returns illustrates this wonderfully. This model began as a 3 factor linear regression that the authors later expanded to 5 terms \cite{famafrench2015}. Currently, the community has recorded over $300$ statistically significant factors for predicting economic returns \cite{cochrane2011discount,harvey2016crosssection}! Random matrix theory says something quite different about that same market. Diagonalizing the return correlation matrix of the S\&P 500 gives a spectrum in which the bulk of eigenvalues falls inside the Mar\v{c}enko--Pastur band expected of pure noise~\cite{marcenko1967distribution}, and the modes that escape it are few and identifiable \cite{laloux1999noise,plerou2002rmt}. The largest is the so-called market mode, whose eigenvector has nearly uniform positive weight on every stock. Below it sit a small number of modes whose eigenvectors localize on industry sectors. Everything else is noise. The Fama-French approach has three structural flaws. Firstly, many of the regressors contain shared information. They may share a few genuine modes that underlie variance but also introduce many spurious dimensions. This is a problem of basis choice. The second is linearity, a bug by construction. If in boom times factors $a$ and $b$ interact in one way and in bust times they behave differently, a linear model will not do (save in trivial cases). Finally, with so many dimensions, we lose any notion of invertibility of the mapping. There could be 10000 combinations of the 500 basis vectors that produce the same result. This prevents the construction of a unique basis of dynamics. 

There have been attempts to pare down these bases, most directly the dynamic factor model, which extracts a small number of common factors from a large panel and is the standard answer to ``many macroeconomic series, few drivers'' \cite{stock2002forecasting}, with established criteria for how many factors to retain \cite{baing2002determining}. Applied to US macroeconomic panels these models routinely settle on four to eight factors \cite{stock2002forecasting,baing2002determining,mccracken2016fredmd}. The factors are linear combinations of the observed series, so the representation is a linear subspace of the raw coordinates and the objection we raised against Fama--French applies unchanged. A linear projection of a nonlinearly folded object mixes genuine modes with spurious ones and cannot recover a curved sheet. A factor model also supplies no metric between two series beyond their loadings, so there is no well-defined distance in which to ask whether two series share driving dynamics.

Large data models, from RMT to transformers, attempt to make up for these shortcomings with data volume, and such high dimensional probabilistic models carry many risks. The rule sets put forward by economists are invariably mangled and lost in the infinite dimensional soup of a transformer world model. Standard macroeconomic techniques, however, such as LASSO based linear regressions have no hope of identifying the complex nonlinear couplings between economic drivers that produce the low dimensional laws in the first place. Analogous limits appear in protein engineering. Direct-coupling analysis fits probabilistic Potts models to multiple sequence alignments and recovers native residue contacts \cite{morcos2011direct,ekeberg2013improved}, but only where the alignment is deep within a single, highly self-similar homologous family, the SH3 domain being the classic covariation case \cite{larson2000sh3}. Such families occupy a vanishingly small and strongly correlated neighborhood of a design space of $20^{{\sim}230}\approx10^{300}$ candidate sequences, so the inferred couplings capture small variations about known homologs rather than modes that span the space, and because sequence covariation is a static observable they cannot recover dynamics at all.

This example bears on economics. There are millions of economic time series to choose from. Walmart alone tracks hundreds of millions of individual products! Tracking each and every variable that constructs a nation's GDP, much less every quantity with some assignation of a property right (clean air, clean water, and so on), escapes tractability. A purely probabilistic world model will always suffer from three issues: 1) A curse of dimensionality. 2) A lack of sampling with regime switching leaving probabilistic models without in-distribution support. 3) In a world of fat tails, a probabilistic model of the universe misses the dynamics that lead to them. This causes systematic failure when these models are most needed! The objective of this work is to provide a universalizable protocol by which we can overcome these gaps up to the informational content fed into the system.

Several recent data-driven-dynamics frameworks share some of this motivation but differ in basis philosophy. Dynamic Mode Decomposition~\cite{schmid2010dmd} and Koopman operator methods~\cite{mezic2005spectral,budisic2012applied} linearize dynamics in a lifted space while we preserve nonlinearity in the geometry of the attractor itself. Sparse Identification of Nonlinear Dynamics (SINDy)~\cite{brunton2016sindy} assumes a sparse symbolic basis of elementary functions while our geometrical approach does not require clean analytical forms in low dimension. Convergent Cross Mapping~\cite{sugihara2012ccm} also uses Takens delays, but for pairwise causality whereas we extend to a global basis via Gromov-Wasserstein transport and a shared barycentric coordinate system. None of these methods produces a shared metric-measure space in which arbitrary timeseries can be co-embedded with a well-defined distance. This is the fundamental contribution of the approach put forward in this paper that makes econometrics tractable for the macroeconomy and provides a universalizable path to a shared and compact basis for timeseries.  

The question that techniques in machine learning such as contrastive learning have failed to address when applying this idea to the general problem of transfer learning is that of well defined metric. The relationship between geometries built from two separate timeseries is not \emph{a priori} defined. Undisciplined approaches attempting to relate embeddings of poorly defined (often not uniform) metric or different well defined metrics lack both mathematical rigor and interpretability. Now, with the advent of approaches like Riemannian-geometric variational autoencoders~\cite{arvanitidis2018latent}, Mahalanobis-distance dimensionality reduction~\cite{singer2008nonlinear}, metric manifold learning \cite{perraultjoncas2013nonlinear} and diffusion maps \cite{coifman2006diffusion}, one can construct latent spaces of well defined metric. However, solving the formal optimal transport problem between two manifolds of defined metric (Gromov--Wasserstein~\cite{memoli2011gromov}) scales poorly ($\mathcal{O}(n^2) $), making it intractable on large data sets.

\subsection{Dynamics as an antidote}

Almost half a century ago, Packard, Crutchfield, Farmer, and Shaw~\cite{packard1980geometry} wrote a seminal paper showing the relation between a single variate timeseries and the geometric attractor that defines its dynamics. The year after, Floris Takens \cite{takens1981detecting} introduced his theory demonstrating that any arbitrary timeseries could be structured in such a way that its dynamics would be entirely encoded in the structure of the manifold that those observations live on. Now it was understood that under certain modest technical conditions, we no longer needed differential equations of closed form to describe the evolution of novel dynamical systems. This could all come from geometry extracted directly from partial observables of the dynamics we care about.

This paper does not present a model for the economy. Rather, it presents a method for producing a basis to describe its dynamics. The world evolves according to low dimensional laws living in high dimensional spaces. A scalar delayed-feedback equation with a handful of parameters is enough to produce output that looks arbitrarily complex~\cite{mackey1977oscillation}, and the task is to recover the former from the latter. Whatever infinite compute and probabilistic models drowning themselves in data can achieve, such an approach is untenable here. The more data streams we have, the higher our dimensionality and the more complex the ultra high parameter space models are to navigate. The Solow model relies on just a handful of factors such as land, labor and capital to explain changes in growth of the macroeconomy \cite{solow1956contribution}. The Phillips curve, relating unemployment to wage growth in its original form and to price inflation in the form used since~\cite{phillips1958relation,samuelson1960analytical}, is expected to span only a few dimensions. And this is to say nothing of the fact that Friedman's velocity of money equation \cite{friedman1956quantity} and the monetary transmission chain suggest much economic activity is parametrized by the price of money, and velocity itself remains partly forecastable from standard macroeconomic aggregates, at least at short horizons \cite{jung2017money}.

In this work, we combine the power of Takens' Delay Embedding Theorem, the rigor of the Gromov--Wasserstein Optimal transport pipeline and the dynamics inherent in $281{,}536$ macroeconomic and financial timeseries drawn from 17 sources, principally Eurostat, the European Central Bank and the World Bank, together with the Penn World Table, the OECD, the Bank of England, the Bank for International Settlements, the International Monetary Fund and the Federal Reserve's FRED archive, and market data providers, to construct the geometry of the macroeconomy in a single, tractable and interpretable manifold, complemented by per-law barycenter architectures for regime detection (\S\ref{sec:prediction}). We dispose of the curse of dimensionality by extracting the low dimensional dynamical basis upon which these phase spaces evolve. By providing that common support, only the orthogonal components of each timeseries contribute to new directions of variation, eliminating this problem of shared support across many regressors. This phase space of phase spaces is a geometric object that itself is nonlinear and therefore permits complex relationships between its supports. However, the origin of each attractor is tractable and decomposable meaning that this phase space of phase spaces is not a black box but rather the geometry of the macroeconomy itself. We demonstrate this by recapitulating diverse results of macroeconomic theory in a compact way on our latent space.

Section~\ref{sec:math} discusses the tools we use to construct this \textbf{Phase Space of Phase Spaces (PSoPS)}. Section~\ref{sec:structure} presents a universal macroeconomic PSoPS drawn from a $\sim$281K-series corpus and its law-scoped re-validation, a geometrical structure defining the relational geometry of individual barycenters that describe conceptual drivers of the macroeconomy. Section~\ref{sec:prediction} shows how the velocity of the transport plans turns the basis into a regime detector, on both the universal PSoPS and the per-law barycenters.

\begin{figure*}[!tp]
\centering
\includegraphics[width=\textwidth]{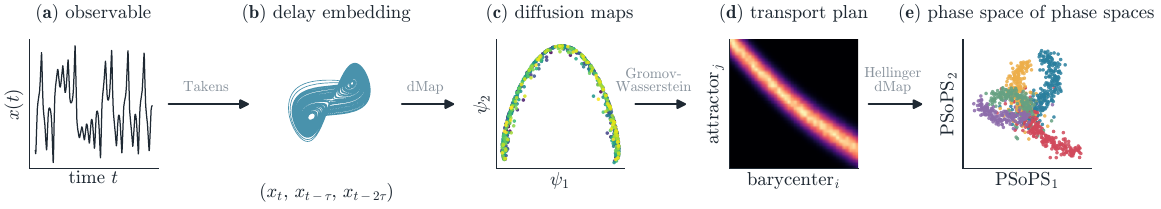}
\caption{\textbf{The PSoPS pipeline.} From scalar timeseries to the Phase Space of Phase Spaces in five stages. Panels correspond one-to-one to \S\ref{subsec:embedding}--\S\ref{sec:gw} and to modules \texttt{aoa\_v3.embed/dmap/gw/barycenter/aoa}. (a) begins with single variate time series observations. Constructing Takens' Delay Embedding vectors brings each time series to a representation of its phase space (b) that is diffeomorphically related to the full dimensional dynamics that generated it. We then apply diffusion maps to reduce the dimensionality of the embedding space further, shown in (c). In (d) we use Gromov--Wasserstein optimal transport to learn the maps between these manifolds extracted from a large number of time series. (e) is the final PSoPS learned from a diffusion map embedding of the family of transport plans between embedding spaces. Color in (c) indexes time along the trajectory, brighter entries in (d) carry more transported mass, and color in (e) distinguishes series families.}
\label{fig:pipeline}
\end{figure*}

\section{Methods and Mathematics}\label{sec:math}

The full construction is summarized in Fig.~\ref{fig:pipeline}. It relies on several key results of dynamical systems theory, namely Takens' Delay Embedding Theorem and Gromov--Wasserstein transport. They describe how we get from the single variate timeseries that dominate economic data to geometric attractors that embed their full dimensional dynamics, then to barycenters that are the Fr\'echet means of related attractors, and finally to the tensor of transport maps that describes the relation between each barycenter. While the individual barycenters give us dynamics, it is this relational geometry that describes how these work together. The computational chain itself, landmark draw to barycenter fit to per-series plans, is laid out as the Architecture list in \S\ref{sec:gw}. 

\subsection{Delay-Coordinate Embedding}
\label{subsec:embedding}

\subsubsection{Takens' Theorem}

Takens' Theorem begins with a single variate time series observation of a dynamical system, for instance the $x$ coordinate of a bob in a chaotic pendulum. Takens' Theorem tells us that if we observe that coordinate for a long enough time such that we well sample its phase space, that is to say it has been ergodically sampled, then we can construct a time delayed representation of its coordinates that contains information about the dynamics driving the system. There also exists a mapping between this space and the full dimensional dynamics produced by that differential equation or series of coupled differential equations that is one to one onto its image (an invertible map between the attractor and its embedded copy). This is wonderfully profound as it means that timeseries, well sampled, are sufficient to resolve what drives any arbitrary dynamical system.

Stated more formally, let $\mathcal{M}$ be a compact smooth $m$-dimensional manifold, $\varphi: \calM \to \calM$ a $C^2$ diffeomorphism, and $h: \mathcal{M} \to \R$ a $C^2$ observation function.

\begin{theorem}[Takens, 1981]
\label{thm:takens}
For generic pairs $(\varphi, h)$, the \emph{delay-coordinate map}
\[
  F_{\varphi,h}^{(d)}(x) = \bigl(h(x), h(\varphi(x)), h(\varphi^2(x)), \ldots, h(\varphi^{d-1}(x))\bigr)
\]
is a diffeomorphism from $\calM$ into $\R^d$ provided $d \ge 2m+1$.
\end{theorem}

The ``generic'' qualifier means that the set of pairs $(\varphi,h)$ for which the embedding fails is nowhere dense in the appropriate function space \citep{takens1981detecting}, with the measure-theoretic strengthening, failure on a shy (prevalence-zero) set, due to \cite{sauer1991embedology}. That is to say we can choose almost any observable function and almost any dynamics and the result holds.

Sauer, Yorke and Casdagli \cite{sauer1991embedology} extended Takens' result to fractal attractors showing that this result holds even in the absence of a smooth mapping. That is to say if $\calA \subset \calM$ has box-counting dimension $d_{\mathrm{box}}(\calA)$, then a generic observation function $h$ still yields an injective delay-coordinate map if $d > 2\,d_{\mathrm{box}}(\calA)$.

Takens' theorem relies upon ergodicity and determinism. We require the following for an exact diffeomorphism: compactness of the manifold $\calM$ (i.e.\ data drawn from a stationary process), determinism of the trajectory and hence a deterministic $C^2$ diffeomorphism (no stochasticity), and exactness of the observation function (no measurement noise). Fortunately, under ergodic bounded stochastic forcing, a generalized (forced) delay embedding remains injective almost surely, at the cost of replacing the diffeomorphism with an injection into a skew-product \cite{stark1999delay,stark2003delay}. Consequently, this method of delays still provides empirical representations of dynamical structure and more importantly shared structures that linear methods miss \cite{Ferg16,Ferg18,ferguson2011cpl,wang2018study}. We have employed this technique in previous work to resolve the dynamical structures of proteins subject to both Gaussian and Poisson distributed measurement noise and stochastic forcing \cite{topel2020,topel2023}.

Using traditional econometric preprocessing we can restore stationarity \cite{hamilton1994timeseries}. In practice each series is routed through a stationarity battery before embedding: ADF and KPSS tests with trend, HEGY seasonal unit-root tests, a GPH fractional-integration estimate and a Bai--Perron break scan decide whether the raw series or its $\tau$-differenced version is embedded, so each attractor is built on $n - d\tau - \tau$ delay vectors when differencing is applied. The battery and its thresholds ship with the code (\texttt{vendor\_atlas/config.py}).

\subsubsection{Delay Parameter Selection}

\textbf{Embedding delay $\tau$:} There are a number of ways of selecting time delays. Given the weekly, monthly, quarterly and yearly granularity of most macroeconomic timeseries, the obvious choice of time delay $\tau$ is given by that temporal granularity. However as different dynamics are accessible on different timescales, it is often desirable to resample say weekly data up to a quarterly level. We use resampling to access attractor dynamics from minute level trading data all the way up to yearly patterns. For arbitrary higher frequency timeseries, we apply the first local minimum of the average mutual information (AMI) method of Fraser and Swinney \cite{fraser1986independent}. In the macroeconomic corpus itself every delay is a calendar ratio set by the reporting frequency ($\tau \in \{1, 3, 4, 5, 12, 21, 63, 252\}$ samples, the calendar and trading-day counts covering week, month, quarter and year), and the AMI rule comes into play only on the synthetic validation systems, where no reporting calendar exists.

For the joint probability estimated via histogram of results, $p(x_i, x_{i+\tau})$, the AMI at lag $\tau$ is
\[
  I(\tau) = \sum_{i,j} p(x_i, x_{j+\tau}) \log \frac{p(x_i, x_{j+\tau})}{p(x_i)\,p(x_{j+\tau})}.
\]
The first local minimum of $I(\tau)$ is then the smallest lag for which observations are as ``independent'' as possible. 

\textbf{Embedding dimensionality $d$:} Takens' Theorem demands that the embedding dimensionality satisfy $d \ge 2m+1$ where $m$ is the intrinsic (manifold) dimensionality of the system (consistent with Theorem~\ref{thm:takens}). We resolve the embedding dimensionality using the false nearest neighbors (FNN) algorithm \cite{kennel1992determining}. A neighbor pair $(i,j)$ in $d$-dimensional embedding space is said to be ``false'' if the $(d+1)$$^{th}$ coordinate sees a gap emerge:

\begin{equation}
  \label{eq:fnn}
  \frac{|x_{d+1}(i) - x_{d+1}(j)|}{\|\dvec_d(i) - \dvec_d(j)\|} > R_{\mathrm{tol}}.
\end{equation}
We set $R_{\mathrm{tol}} = 15$ and a second criterion $\|\dvec_{d+1}(i) - \dvec_{d+1}(j)\|/\sigma_s > A_{\mathrm{tol}} = 2$ to handle noise. However, as data is often noisy, the FNN fraction never asymptotes at zero. The relationship between the residual FNN fraction and the noise itself is given by $(\varepsilon/\sigma_s)^2$ where $\varepsilon$ is the noise level and $\sigma_s$ the signal standard deviation. We set convergence to residual FNN fraction $\leq \eta = 0.01$. When the residual fraction never falls below $\eta$, which is the typical case for noisy macro series, the embedding falls back to the default $d=8$, and $d$ is bounded to $[3,16]$ throughout.

\subsubsection{Global Normalization}

Economic timeseries like GDP grow a lot over time. The regular relationship between say capital, land, labor and GDP observed in the Solow model describes a rate of growth visible in differencing. However the exponential nature of that growth is the impetus for the log scale introduced in Solow. Here we must follow a similar logic for such unbounded timeseries. Each delay vector $\dvec(t) \in \R^d$ is row-normalized:
\[
  \hat{\dvec}(t) = \frac{\dvec(t) - \mu(t)}{\max(\sigma(t),\; \varepsilon)},
\]
where $\mu(t)$ and $\sigma(t)$ are the mean and standard deviation across the $d$ components, and $\varepsilon = 10^{-8}$ prevents division by zero. Using an affine (smooth and monotone) transformation means that we can still apply dynamical classifiers like permutation entropy that are invariant under such transformations.

\subsection{Per-Series Diffusion Maps}
\label{subsec:dmap}

Now that we have recovered the dynamics of each timeseries via Takens' Theorem, we apply a layer of dimensionality reduction and denoising to extract a $d$ dimensional representation of dynamics of each timeseries in a space with defined metric. Diffusion maps work by constructing a random walk on a weighted graph whose weights are extracted from the similarity of any two point cloud estimates. The mathematics is laid out step by step below, following Coifman and Lafon~\cite{coifman2006diffusion}.

Our Takens' Embeddings are point clouds $\{x_i\}_{i=1}^N \subset \R^{d}$, with $d$ the embedding dimension chosen above. 

\textbf{Step 1: Kernel Construction} We begin by applying a Gaussian kernel with adaptive bandwidth adjusted for varying density across the attractor:
\[
  K_0(x_i, x_j) = \exp\!\left(-\frac{\|x_i - x_j\|^2}{\sigma(x_i)\,\sigma(x_j)}\right),
\]
with $\sigma(x_i)$ being the distance from $x_i$ to its $k$-th nearest neighbor ($k=10$ by default). A Theiler window of width $\tau$ excludes temporally adjacent delay vectors from the kernel (entries at $|i-j|\le\tau$ are zeroed), so that kernel neighbors reflect recurrence on the attractor rather than adjacency in time \cite{theiler1986spurious}.

\textbf{Step 2: Non uniform sampling density} We are not guaranteed uniform sampling density. The $\alpha$ parameter can be tuned to compensate. Consequently we define $d_\alpha(x_i) = \sum_j K_0(x_i, x_j)$ and set
\[
  K_\alpha(x_i, x_j) = \frac{K_0(x_i, x_j)}{d_\alpha(x_i)^\alpha \, d_\alpha(x_j)^\alpha}.
\]
With $\alpha = 1$, after the subsequent row-normalization, the resulting Markov generator $(P-I)/\varepsilon$ converges as $N\to\infty$, $\varepsilon\to 0$ to the Laplace--Beltrami operator on the underlying manifold, independent of sampling density \cite{coifman2006diffusion}. For $\alpha<1$ the limit is a density-weighted operator.

\textbf{Step 3: Markov matrix} The probability of jumping from any one state to any other must sum to 1 on every row of the transition matrix (row-stochastic) by conservation of probability mass. We must then row-normalize: $P(x_i, x_j) = K_\alpha(x_i, x_j) / \sum_k K_\alpha(x_i, x_k)$.

\textbf{Step 4: Eigendecomposition} $P$ has eigenvalues $1 = \lambda_0 \ge \lambda_1 \ge \lambda_2 \ge \cdots$ and right eigenvectors $\psi_k$. The \emph{diffusion coordinates} are $\Phi_t(x_i) = (\lambda_1^t \psi_1(x_i), \ldots, \lambda_r^t \psi_r(x_i))$, where $r$ is the number of retained components.

The diffusion distance or distance on the diffusion map landscape is given by 
\begin{flalign*}
  D_t^2(x_i, x_j) &= \sum_{k \ge 1} \lambda_k^{2t}\bigl(\psi_k(x_i) - \psi_k(x_j)\bigr)^2 \\
  &= \|\Phi_t(x_i) - \Phi_t(x_j)\|^2 &&
\end{flalign*}
It is a well-defined metric for any graph Laplacian and the graph Laplacian converges as $N \to \infty$ with bandwidth $\varepsilon \to 0$, to the (weighted) Laplace--Beltrami operator on a smooth manifold. 

This spectral decomposition provides us with a distribution of eigenvalues. In the corpus pipeline each attractor is first reduced to its $n_s = 40$ support points by Farthest Point Sampling (FPS), the greedy rule that repeatedly picks the point farthest from those already chosen so that the supports cover the attractor rather than its densest fold and the diffusion map is computed densely on that support, so the per-series eigendecomposition is always small. Every series leaves this stage as the same kind of object, a $40\times 40$ diffusion distance matrix on its support points normalized to unit maximum, and it is this matrix that serves as the cost matrix in every transport step below. Dimensionality itself is estimated at the corpus level, where the Coifman scaling estimator is paired with the L-method knee of Salvador and Chan \cite{salvador2004determining} (\S\ref{subsec:universal-aoa}).

\section{Gromov--Wasserstein Alignment}
\label{sec:gw}

At this point we have taken in our timeseries and computed from them geometrical descriptions of each in latent space. Relating them to one another is the core role of Gromov--Wasserstein Optimal Transport (GW-OT). The idea is that variation in distance on the manifolds $\calX$ and $\calY$, both of which have their own distinct but well defined metrics can be used to learn the optimal map between points on $\calX$ and those on $\calY$. The idea that proximal points share similar dynamical information is inherent in dMaps. Now the closeness of the dynamical structures revealed in $\calX$ and $\calY$ can be understood in relation to each other. 

\subsection{GW Distance}
\label{subsec:gw-distance}
The Gromov--Wasserstein distance compares metric measure spaces that share no ambient coordinate system, which is exactly the situation delay embeddings leave us in~\cite{memoli2011gromov}.

\begin{definition}
Let $(\calX, d_{\calX}, \mu_{\calX})$ and $(\calY, d_{\calY}, \mu_{\calY})$ be two metric measure spaces.
The \emph{Gromov--Wasserstein distance} is
\[
\begin{aligned}
\dGW^2(\calX, \calY) ={}&
\inf_{\gamma \in \Pi(\mu_\calX, \mu_\calY)}
\iint\! |d_\calX(x,x') - d_\calY(y,y')|^2 \\
&\,\mathrm{d}\gamma(x,y)\,\mathrm{d}\gamma(x',y')
\end{aligned}
\]
where $\Pi(\mu_\calX, \mu_\calY)$ is the set of couplings with marginals $\mu_\calX$ and $\mu_\calY$.
\end{definition}

The difference in the distances between pairs $x$ and $x^{\prime}$ and $y$ and $y^{\prime}$ can be used to measure how much the distance between a pair in $\calX$ and its counterpart pair in $\calY$ varies. $\gamma$ is a measure of the coupling between points in $\calX$ and points in $\calY$. It is a joint distribution over $\calX \times\calY$ with marginals equal to the measures on $\calX$ and $\calY$ and can be thought of as a transport plan between points on the two. Integrating over all possible pair choices and minimizing these couplings over all possible transport plans, we get the Gromov-Wasserstein distance, $\dGW$. Note that $\calX$ and $\calY$ can live in completely different spaces because we are only comparing distances between points in the point clouds.

\subsection{Entropic Regularization}

GW optimization is not convex and consequently we need some kind of regularization \citep{peyre2016gromov}. This is because $\gamma$ appears twice in the integrand so we get a quadratic form in $\gamma$. As $\gamma$ is a coupling matrix over a finite support, the problem reduces to a (non-convex) Quadratic Assignment Problem (QAP-hard) with multiple local minima. To prevent getting stuck in the ``wrong'' minimum we have to introduce regularization in any gradient descent. This is done by adding some entropic term with strict concavity ($H(\gamma)=-\int \gamma \log\gamma$) thereby smoothing out sharp local minima. It also forces $\gamma$ to be strictly positive everywhere, eliminating sparsity, and this strict positivity is what the later operations with $\gamma$ require.

\begin{flalign*}
  \dGW^{2,\varepsilon}(\calX, \calY) &= \inf_{\gamma \in \Pi} \left\{ \iint L \,\mathrm{d}\gamma\,\mathrm{d}\gamma + \varepsilon \, H(\gamma) \right\} &&
\end{flalign*}
where $L(x,x',y,y') = |d_\calX(x,x') - d_\calY(y,y')|^2$ and $H(\gamma) = -\int \gamma \log \gamma$ is the entropy.

This is solved using projected gradient descent with Sinkhorn iterations at each step \cite{cuturi2013sinkhorn}. The Sinkhorn algorithm solves the entropy-regularized linear OT subproblem obtained at each mirror-descent step when the GW quadratic form is linearized about the current $\gamma$ \cite{peyre2016gromov}. The value of $\varepsilon$ must be tuned in order to not have an entropy term so small that we still have to deal with a lack of convexity causing Sinkhorn to not converge nor too large that entropy dominates (uniform matrix where all transport plans are the same). In practice this optimal value is derived empirically to ensure we do not recover degenerate states but are able to navigate the non-convex quadratic optimization surface. We also find empirically that the optimal $\varepsilon$ is ${\approx}0.008$ regardless of training corpus size, tested from 1000 to 20000 timeseries. Note optimality here is defined by consistency in implied dimensionality across batched randomized training data across variation in $\varepsilon$ and number of supports, $n_s$. 

\subsection{GW Barycenter: The Phase Space of Phase Spaces}

GW becomes increasingly intractable for the entirety of the time series space as we would need transport maps between every attractor pair which grows like $n^2$. Instead we compute some number of reference frames, so-called barycenters. For the universal corpus of this paper a single barycenter turns out to suffice (\S\ref{subsec:universal-aoa}), and nothing in the construction changes if several are used. The barycenter is a metric measure space that is to say a set of points with a well defined distance function (metric) and a measure (weights) determining the mass or importance of each point. 

\begin{definition}
Given metric measure spaces $\{(\calX_i, d_i, \mu_i)\}_{i=1}^K$ with weights $\{\lambda_i\}$, the \emph{GW barycenter} is
\[
  \calB^* = \arg\min_{\calB} \sum_{i=1}^K \lambda_i \, \dGW^2(\calB, \calX_i).
\]
\end{definition}

This is just the Fr\'echet mean in GW space. This quantity is well defined in any space with a distance function so now it is up to us to find that distance function. So the Fr\'echet mean is the optimal barycenter that is simultaneously closest in GW distance to all landmark attractors from which it is constructed. This is a lossy summary but the lost information remains in the transport plan. 

So now we have two optimizations to do. The first being the $d_{GW}$ and the second being the $\calB^*$ calculation. We do this in two steps. First we fix the barycenter $\calB$ and optimize all $\gamma$. Then we fix $\gamma$ and optimize $\calB$. We repeat until convergence.

In practice the barycenter is composed of $n_s=40$ barycenter points with equal weight, initialized with a random distance matrix. It is fitted to a pool of $1{,}500$ landmark series drawn by a seeded concept-balanced rule, at most 120 series per concept, truncated to $1{,}500$ in concept order so that 15 of the 22 concept groups enter the fitted pool. The reference therefore reflects diverse macro structure rather than the most-published families. This fitted pool is distinct from the 800 landmarks used later to carry the diffusion-map eigenbasis (\S\ref{subsec:universal-aoa}). The two serve different steps and need not match. Each landmark series contributes its diffusion distance matrix $D_i$, restricted to the 40 representative points chosen by FPS on its own attractor and normalized to unit maximum so that no series dominates by scale. At 1500 of $17{,}025$ series the balanced draw samples the corpus geometry densely, which is why the Fr\'echet mean is insensitive to it (Fig.~\ref{fig:aoa-robustness}A(i)). Now we want to produce a single distance matrix $D_B$ that describes the average geometry or barycenter. 

We solve the exact GW problem between the current $D_B$ and each landmark to find the $\gamma$ that best preserves structure. All of any given landmark point's mass has to go somewhere and the sum of all mass at all barycenter points must match that given by the landmark points (marginal constraints). The reference geometry is fitted without entropic regularization. Entropy enters only afterwards, in the per-series plans onto the frozen barycenter, where the soft matching it induces is what spreads each point's mass over several barycenter points probabilistically.

We now have 1500 transport plans (one per landmark), each a $40\times 40$ doubly stochastic matrix (entries positive, rows and columns summing to $1/n_s$). We update $D_B$ with $\gamma_i^\top D_i \gamma_i$ which is effectively the average distance in landmark $i$'s space between the mass that maps to barycenter points $b$ and $b'$. In doing so we iteratively reshape the distance matrix between the points until we converge on an averaged structure.

\textbf{Architecture:}

Rather than computing $O(n^2)$ pairwise GW distances among all $n$ series, we:
\begin{enumerate}
  \item Draw the concept-balanced landmark pool, at most 120 series per concept, truncated to $1{,}500$ in concept order.
  \item Compute the GW barycenter from these landmarks.
  \item Project each of the $n$ series individually onto the barycenter.
\end{enumerate}

The per-series entropic-GW solve against a frozen barycenter costs $\sim0.5$\,s on one core at $n_s=40$, so projecting a corpus of $\sim$$10^5$ series is a few hours on eight cores rather than the minutes a naive estimate suggests. The barycenter solve itself is $\sim$3\,s and the landmark selection negligible. Timings are single-core on an Apple M-series CPU. The full corpus extension of \S\ref{subsec:extension} took $10.2$ hours across eight workers. 
 This is the workflow used for any extension to the full 281k corpus. Per-series GW is run for every series, while a dense Hellinger-dMap over all plans (the diffusion map on transport plans constructed below) would be $O(N^2)$ and intractable at 281K series. We build the diffusion-map eigenbasis on 800 FPS landmark plans and place every other series by Nystr\"om-extending that eigenbasis with the series' own plan, avoiding the dense $N^2$ spectral step, not the per-series GW. The extension requires that the landmarks cover the plan geometry and that the entropic regime give graded plan similarity rather than near-orthogonal plans.

We have now constructed the barycenter. However, the barycenter itself is not the desired object as its averaged structure does not represent any one phase space well. It is the transport plans themselves that carry all of the information about the original attractors. For series $i$, we compute the entropic GW transport plan $T_i \in \R^{n_s \times n_s}$ between its attractor and the barycenter. We write $T$ for a converged plan and reserve $\gamma$ for the coupling while it is still the variable of an optimization. $T_i$ is a doubly stochastic matrix with uniform marginals: $\sum_j T_i(a,j) = \sum_a T_i(a,j) = 1/n_s$. The entry $T_i(a,b)$ is then the fraction of attractor point $a$'s mass sent to barycenter point $b$.

This coordinate system embeds all of the information and if the main drivers of time series $i$ and timeseries $j$ are the same they should map to the same parts of the generalized attractor. However, if we have two very similar objects like French and German GDP, the distances between the two transport maps may not be sufficiently big to be well resolved. If for instance, we look only at Polish voivodeship level GDP data, we may end up with transport plans near the uniform coupling ($U = \frac{1}{n_s^2}\mathbf{1}$). To combat this, we use the Hellinger distance metric which amplifies differences by a factor of $\sim n_s/2$ relative to Frobenius (a $20\times$ signal boost at the $n_s = 40$ used here). 

The set of all $n_s\times n_s$ doubly stochastic matrices is known as the Birkhoff polytope, written $\mathcal{B}_{n_s}$ below and not to be confused with the barycenter $\calB$. Its vertices are the $n_s!$ permutation matrices, and the polytope itself has dimension $(n_s-1)^2=1521$ for $n_s=40$. We use the full transport-plan matrix $T_i \in \mathcal{B}_{n_s}$, not a row-averaged summary. While this is large, it is far smaller than the $17{,}025$ transport plans used to construct this system. However, as the attractors themselves are strongly coupled and low dimensional, we do not need the entire 1521 dimensions. Rather, we need some low dimensional submanifold within it. One can imagine that if all $n_s=40$ supports carried equal mass, then attractor $i$, carries no information about attractor $j$. Conversely, if all of the mass in some eigenvector of attractor $i$ maps uniquely to an eigenvector of attractor $j$, most of the Birkhoff polytope vertices are unnecessary. The true solution is somewhere in the middle and is muddied by entropic regularization. To recover this low dimensional submanifold of the Birkhoff polytope on which the plans live, we apply diffusion maps once again, this time with Hellinger distances for our distance matrix. The square root inside the Hellinger metric is what buys the amplification, stretching differences between small transport masses and compressing those between large ones, so plans that differ only in fine structure become resolvable. We describe the algorithm briefly here: 

\begin{enumerate}
  \item Each $T_i$ is mapped to $\sqrt{T_i}$ (element-wise square root), embedding $\mathcal{B}_{n_s}$ into a submanifold of the unit sphere.
  \item Pairwise distances are computed in the Hellinger metric: $d_H(T_i, T_j) = \|\sqrt{T_i} - \sqrt{T_j}\|_F$.
  \item A Gaussian kernel at a single global bandwidth, the median pairwise Hellinger distance, is constructed on these distances.
  \item Alpha-normalization ($\alpha = 1$) gives the Laplace--Beltrami operator on the transport-plan manifold.
  \item Eigendecomposition yields PSoPS diffusion coordinates: the first $d$ eigenvectors of the diffusion operator on $\mathcal{B}_{n_s}$.
\end{enumerate}

We need a sufficient $n_s$ to resolve differences in attractors. Empirically, $n_s=40$ was optimal for the $17{,}025$-series global corpus. At smaller levels near $n_s=10$ the transport plan is too coarse and fine grain structure is eliminated. At large $n_s$, say 200, entropic regularization pushes us toward a uniform matrix. Naturally there is then an interplay between the entropic regularization parametrized by $\varepsilon$ and the choice of $n_s$. At the empirically optimized $\varepsilon=0.008$ we observe a phase transition near $n_s\sim35$, where the spectral gap of the Hellinger DMAP eigenvalue spectrum jumps from $1.37$ to $2.39$. Bootstrapping across 10 replicates of 80 series, $n_s=40$ holds a strong gap of $2.12\pm0.28$ with the lowest coefficient of variation of any stable choice, so it is the value used throughout.

To summarize, each series' attractor has a shape and the transport plan to the barycenter is a near-lossless encoding of that shape as it relates to a common geometric object, subject to the entropic-regularization resolution limit parametrized by $(n_s,\varepsilon)$. Similar attractors will yield similar encoding and hence be proximal in the phase space of phase spaces and vice versa. The Hellinger geometry of the transport plans on the universal macro corpus yields an effective dimension $d_{\mathrm{eff}} = 7.3 \pm 0.2$ (Coifman $\varepsilon$-scaling~\cite{coifman2006diffusion}, see \S\ref{sec:results}, Table~\ref{tab:deff}), measured on all $17{,}025$ series with no resampling.

We estimate dimensionality using Coifman $\varepsilon$-scaling dimension, $d = 2\max_{\varepsilon} \mathrm{d}\log S(\varepsilon)/\mathrm{d}\log\varepsilon$ with $S(\varepsilon)=\sum_{ij}\exp(-\|x_i-x_j\|^2/\varepsilon)$ chosen for its insensitivity to the sampling density that troubles short-range distance statistics like nearest-neighbor maximum-likelihood estimators. By determining dimensionality from the steepest point of a log-log curve, no bandwidth is chosen and none can be tuned. We pair this with the L-method knee~\cite{salvador2004determining}. The two estimators, while consistent within uncertainty, differ in precision rather than in answer, for structural reasons. The L-method locates the knee as the intersection of two straight lines fitted to the eigenvalue curve that minimize the total fitting error. This intersection responds nonlinearly to perturbation as a small change in the curvature of the spectrum can move the crossing point by several indices. The Coifman dimension reads a slope rather than a crossing and inherits the spectrum's variance directly (smooth rather than discrete).  


\subsection{Structure}
\subsubsection{Building a unified basis}

Takens' Theorem, while powerful, introduces several major challenges. Firstly, each embedding we make lives in its own space. While each has \emph{a} well defined metric, the comparison between these two is \emph{a priori} unknown. We solve this issue by considering the Gromov--Wasserstein optimal transport between each manifold. This, however, becomes problematic as the number of GW solves scales like the square of the number of timeseries we have (it is a pairwise comparison) and it also fails to provide a common basis. These problems are solved by constructing local barycentric attractors, $B^{*}_i$, that describe the dynamics of groups of similar attractors. The pairwise relations between this much more pared down set of dynamical variables become tractable as we do not need to compute the full pairwise network of transport plans but only local barycenters and the transport plans between them. If Takens provides us with the geometry of timeseries and Gromov-Wasserstein with the relationship between them, then it is these $B^{*}_i$ that embed dynamics on a conceptual level and the transport plans between them that describe how their dynamical modes are related.

A key insight of complex systems theory and indeed dynamical systems is separability of dynamics over different scales. There will naturally emerge a hierarchy of dynamics both temporally and geographically. The optimal set of $B^{*}_i$ at say $\tau=$ 1 month vs 12 months or for global GDP versus UK NUTS data will be different. While all of these dynamics live in a ``Phase Space of Phase Spaces'', it is both computationally and practically convenient to restrict ourselves to a construction of $B^{*}_i$ relevant to the task at hand. 

Secondly, Takens' Theorem assumes ergodicity while economic systems evolve. A key innovation of this work is to track the velocity of the barycenter coordinates and transport plans. Of course the attractors will evolve and even bifurcate with regime switching events. No model can predict dynamics that remain unseen. However, this serves as a regime detection mechanism and suggests when the relevant basis of dynamics changes. The simplest example here is that of the boom-bust cycle. This is perhaps the most well recapitulated effect in econometrics. Boom-bust cycles have been around for as long as economies have been. The relationship between say labor, capital and land changes based on where we are in this cycle and as such the relevant modes of dynamics will as well. 

\subsection{Relational objects: attractors, barycenters, and transport}

We have already dealt with the construction of the individual attractors $\mathcal{M}_i$ and barycenters $B^{*}_i$. These are the attractor geometries themselves. To understand coupling, however, we must understand the relational structure between them: given the factors driving $B^{*}_i$, how predictive are they of $B^{*}_j$?

Let $\{\Phi^i_k\}$ be the set of (normal) slow modes of $\mathcal{M}_i$ and $\{\Phi^j_l\}$ be those for $\mathcal{M}_j$. These are the diffusion eigenvectors $\psi_k$ of \S\ref{subsec:dmap}, written $\Phi^i_k$ so that the series and the mode each carry an index. Let $\{\mathcal{M}_i\}$ be the set of all dynamical manifolds (either $B^{*}_i$ for the coarse graining or simply the attractors extracted from each timeseries). We have already described the GW process that brings us from $\mathcal{M}_i$ to $\mathcal{M}_j$ via some transport plan $T_{ij}$. If $\mathcal{M}_i$ and $\mathcal{M}_j$ describe the same dynamics, then $T_{ij}$ should describe transportation of most mass in single entries. That is to say $T_{ij}$ concentrates near a permutation, diagonal up to an $O(\varepsilon)$ off-diagonal floor set by the entropic regularization (not to be confused with the finite-sample velocity-CV floor discussed in \S\ref{sec:velocity}). If however, $\mathcal{M}_i$ and $\mathcal{M}_j$ share little information, then the mass transported from $\mathcal{M}_i$ should get distributed uniformly over the supports of $\mathcal{M}_j$. This entropy of the transport plan, $H(T_{ij})$ is a scalar proxy for the far richer tensor below, which resolves how each mode of $\mathcal{M}_i$ couples to each mode of $\mathcal{M}_j$ rather than collapsing the coupling to a single number.

It then follows that if we want to understand how much $\{\Phi^i_k\}$ can explain $\{\Phi^j_l\}$, we simply take the dot product of $\{\Phi^i_k\}$ mapped to $\mathcal{M}_j$'s space.

\begin{equation}\label{eq:sensitivity}
   {\mathcal{S}_{i\rightarrow j}^{k,l} \;=\; w^j_l \,\bigl\langle T_{ij}\,\tilde\Phi^i_k,\; \tilde\Phi^j_l \bigr\rangle, \qquad w^j_l \;=\; \lambda^j_l \Big/ \textstyle\sum_{m\ge 1}\lambda^j_m }
\end{equation}

where $\tilde\Phi^i_k \in \R^{n_s}$ is the pushforward of the DMAP eigenvector $\Phi^i_k$ onto the shared $n_s$-point barycenter skeleton $\mathcal{B}^*$ via the per-series plan $T_i: \mathcal{M}_i \to \mathcal{B}^*$ (that is, $\tilde\Phi^i_k = n_s\, T_i^\top \Phi^i_k$ under the uniform marginal $1/n_s$). The effective inter-manifold transport acting on $\R^{n_s}$ is then $T_{ij} = T_j T_i^\top$. The sum in the denominator runs over retained modes on $\mathcal{M}_j$ and $\langle\cdot,\cdot\rangle$ denotes the inner product under the uniform skeleton measure $1/n_s$.

Here $\mathcal{S}_{i\rightarrow j}^{k,l}$ is then the variance in each direction $\Phi^j_l$ explained by each $\Phi^i_k$ given the transport plan $T_{ij}$, weighted by the relative importance of each $\Phi^j_l$ determined by the portion of variance described by its associated eigenvalue.

We can consider some tensor, $\alpha$, with entries $\mathcal{S}_{i\rightarrow j}^{k,l}$ that describes the complete set of what portions of $\mathcal{M}_j$ are described by $\{\Phi^i_k\}$. It is, of course, quite expensive to compute, but is manageable with the barycentric formulation of the phase space of phase spaces. It is a richer version of what $d_{GW}$ coarse-grains and can be understood at arbitrary levels of coarse graining.

Although $T_{ij}$ is a doubly stochastic coupling, $\mathcal{S}_{i\rightarrow j}^{k,l}\neq\mathcal{S}_{j\rightarrow i}^{l,k}$. This is because the features $\{\Phi^i_k\}$ may describe variance of some downstream dynamics $\mathcal{M}_j$ but the reverse may not be true. Consider for instance the phase space of U.S. interest rates versus default rates of personal loans in Canada. U.S. rates are a major driver of global rates and consequently defaults so a primary mode of the latter time series must come from the former. The reverse however is not true. This can also be used for understanding lead--lag relations and formalizing regime changes. The primary modes of a lagging indicator should be described by those of a leading indicator but not vice versa. Similarly for regime change, the velocity of transport plan entries should indicate a change in the structural relationship between any two manifolds and consequently what leads what. 

A symmetric summary of the network of barycenters $B^{*}_i$ could be embedded deterministically via classical multidimensional scaling \cite{torgerson1952mds}, but a symmetric distance is precisely what we do not want here. As coupling between macroeconomic concepts is so strong, attractors constructed to describe ``unemployment'', ``inflation'' and ``GDP'' look like they are quite similar. The asymmetry $\mathcal{S}_{i\rightarrow j}\neq\mathcal{S}_{j\rightarrow i}$ is why we keep directed transport plans, not a symmetric distance. This means that they carry the lead--lag structure a metric discards.

\section{Results}\label{sec:results}

\subsection{Validation on known dynamical systems}\label{sec:validation}

\begin{figure*}[!tp]
\centering
\includegraphics[width=\textwidth]{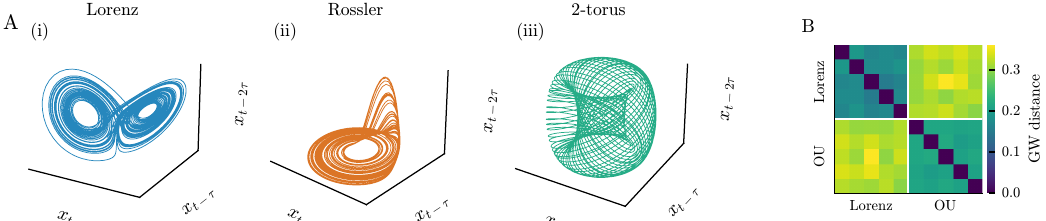}
\caption{\textbf{Validation on known dynamical systems.} \textbf{A}: delay-coordinate attractors of (i)~Lorenz, (ii)~R\"ossler and (iii)~a 2-torus. \textbf{B}: Gromov--Wasserstein distances among five chaotic (Lorenz) and five linear-Gaussian (Ornstein--Uhlenbeck) realizations. The block structure is the result: cross-family distance is $1.6\times$ the within-family median, so the cost geometry alone separates determinism from a matched stochastic null. }
\label{fig:validation}
\end{figure*}

Before applying the pipeline to economic data, we validate it on dynamical systems whose invariants are known. Using scalar observables of the Lorenz and R\"ossler attractors and of a two-frequency quasiperiodic torus we construct a shared barycenter from the delay-embedded, diffusion-mapped, and transported time series observations (Fig.~\ref{fig:validation}A). The recovered geometries reproduce the textbook invariants of each system, confirming that the construction preserves the dynamics it encodes.

 The validation trajectories use standard parameters for each system: Lorenz $(\sigma,\rho,\beta)=(10,28,8/3)$~\cite{lorenz1963deterministic}, R\"ossler $(a,b,c)=(0.2,0.2,5.7)$~\cite{rossler1976equation}, and a two-frequency torus with incommensurate frequencies $\omega=(1,\sqrt{2})$. Each is integrated past its transient and observed through a single scalar coordinate before delay embedding, with $\tau$ at the first minimum of the average mutual information, the embedding dimension selected by false nearest neighbors, and a Theiler window excluding temporally adjacent pairs from the correlation integral~\cite{theiler1986spurious}. The chaos-versus-noise corpus pairs five Lorenz realizations against five Ornstein--Uhlenbeck ones ($\theta=0.5$). All integration settings and seeds live in \texttt{scripts/fig2\_validation.py}. The comparison of record is the delay reconstruction against the true three-dimensional state under the same estimator, the Grassberger--Procaccia correlation dimension $D_2$~\cite{grassberger1983characterization}, which isolates what Takens promises, that reconstruction preserves the invariant, from estimator convention. The reconstruction returns $D_2=2.09\pm0.02$ for the Lorenz attractor against $2.06\pm0.02$ on its true state (literature $2.05$), and $1.89\pm0.02$ for the R\"ossler against $1.87\pm0.01$ on its true state (literature $\approx2.0$). The single scalar observable preserves the invariant to within the estimator's own uncertainty, together with the loop topology of the torus, whose first Betti number, the count of persistent loops, is recovered as $\beta_1=2$. On a corpus of five chaotic (Lorenz) and five linear-Gaussian (Ornstein--Uhlenbeck) realizations, the Gromov-Wasserstein geometry separates nonlinear structure from stochastic nulls, with cross-group distances $1.6\times$ the within-group median (Fig.~\ref{fig:validation}B).
 
The $1.6\times$ figure is the ratio of the median cross-family GW distance to the median within-family distance, so determinism is separated from a matched stochastic null by the cost geometry alone. The choice of the Hellinger metric on the transport plans is carried by the resolution argument of \S\ref{sec:gw}, the ${\sim}n_s/2$ amplification over a Frobenius distance, and it is validated on the macroeconomic corpus itself, where Hellinger coordinates keep $16{,}756$ of $17{,}025$ series distinct at a matched threshold that collapses the majority of raw-plan coordinates into near-duplicates. The pipeline recovers the textbook invariants of known attractors, and the Hellinger geometry is what gives the basis its resolution.

Discrimination across this zoo of attractors is family dependent. To show that delay-embedded geometry distinguishes dynamics beyond a single chaos-versus-noise contrast, we cluster twelve dynamical families spanning chaos, maps, quasiperiodicity, and noise on attractor-geometry and temporal features of the same delay embeddings. These include Lorenz at two parameter sets, R\"ossler chaotic and near-periodic, H\'enon, logistic at two rates, AR(1) at two persistences, random walk, white noise, and a two-frequency torus with five initial-condition draws each. The adjusted Rand index (ARI) scores how well a recovered clustering matches the ground-truth family labels, corrected for chance, with $\mathrm{ARI}=1$ perfect agreement and $\mathrm{ARI}\approx0$ the chance baseline~\cite{hubert1985comparing}. The families separate cleanly with adjusted Rand index $0.88$ against ground truth on the temporal feature set, recovering distinct dynamical classes from single scalar observables. Finally, the per-series diffusion map functions as a denoiser of the separate attractor-geometry feature, a weaker classifier than the temporal set on its own. It improves that feature in both regimes but is essential on the noisy macro data, where the raw delay-coordinate geometry alone carries almost no class signal ($\mathrm{ARI}\,0.00\to0.10$), and merely helpful on the clean simulated systems ($0.23\to0.30$). Clean systems are therefore best served by raw Takens geometry and noisy ones by the diffusion-map denoising.

\subsection{The Topology of the Phillips Curve}
\label{subsec:phillips-loop}

 The Phillips curve \cite{phillips1958relation}, in the price-inflation form of Samuelson and Solow~\cite{samuelson1960analytical}, is a canonical relationship in the macroeconomics literature connecting unemployment to inflation. The econometric argument has largely centered itself along the slope of this relationship. That is to say, economists focus on the sharpening or flattening of this trade-off \cite{blanchard2016phillips,hazell2022slope}, how expectations enter \cite{galigertler1999}, and where the non-accelerating-inflation rate sits. Those are questions about coefficients fit to a linearized model. The result of this section is that the Phillips relation is not a curve but a loop, that the loop is a measurable topological invariant of the joint attractor, and that unemployment leads inflation in the geometry.

This work provides a geometric representation of the Phillips manifold relationship. The claim we make here is of a different kind and, to our knowledge, new. The relation has a topology and this topology is deeply connected to that of the macroeconomy as a whole. If unemployment and inflation traced a curve, however shifting, the joint cloud would be contractible and its first homology would vanish. It does not! The joint attractor, built from monthly FRED data (UNRATE and year-over-year CPI inflation, 1948--2026, $936$ overlapping months), carries persistent one-cycles, which is the geometric signature of hysteresis, and a loop cannot be produced by any single-valued function of unemployment onto inflation regardless of how that function is allowed to drift.

This reframes a decades-old debate about the slope of a curve as a statement about the shape of an attractor, and it is measurable rather than inferred. The theory suggests that when unemployment is low, spending rises, lifting the velocity of money and with it inflation. The converse drives the downswing, and the two together trace the so-called Phillips curve. In this work, we postulate (a) that the timeseries of both unemployment and inflation have corresponding attractors of higher dimension than their univariate timeseries and (b) that it is the relationship between these manifolds embedding their dynamics that is the quantity of interest. Applying the same framework as we did to the toy systems above, we expect that if this coupling between the two variables is strong then we will observe structurally similar attractors with similar transport maps to the barycenter. 

\begin{figure*}[!tp]
\centering
\includegraphics[width=\textwidth]{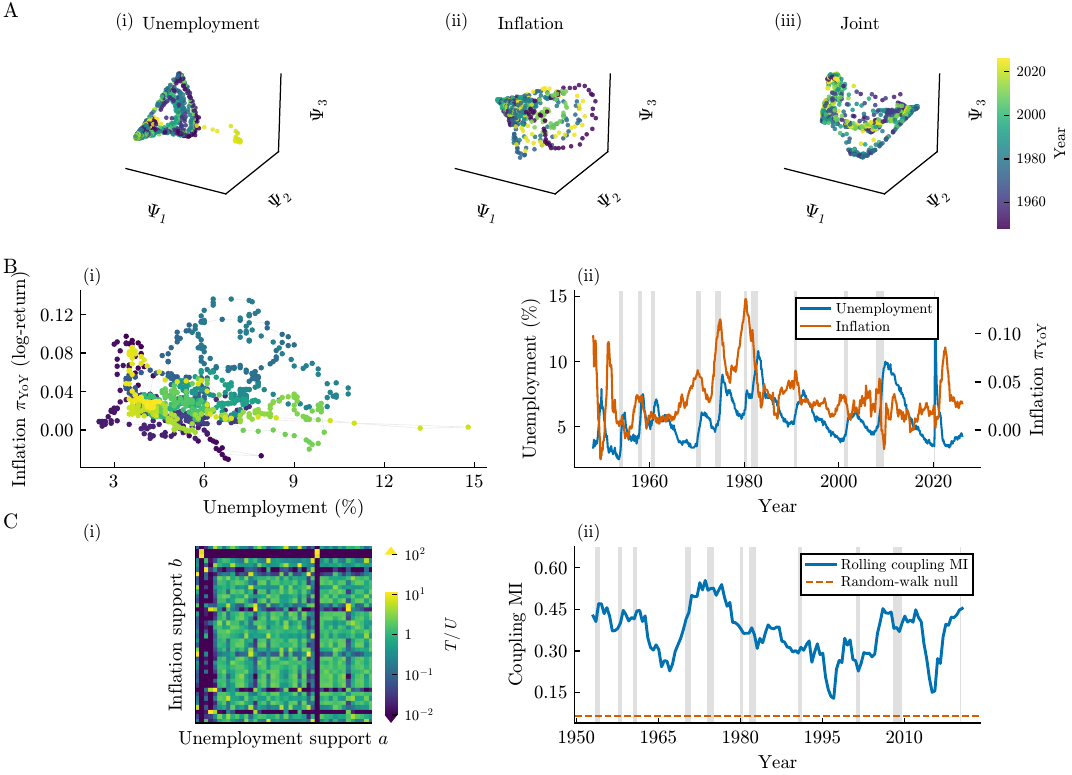}
\caption{\textbf{Unemployment and inflation share a coupled geometry.} \textbf{A}: three-dimensional diffusion-map attractors of (i) unemployment, (ii)~inflation, and (iii) their joint embedding, colored by year. Each carries a persistent loop (first Betti numbers $\beta_1 = 43$, $54$, $39$ for (i)--(iii)). \textbf{B}: the same data in raw form. (i) the Phillips loop in the (unemployment, $\pi_{\mathrm{YoY}}$) plane, colored by year as in A and threaded by a faint gray path in time order, and (ii) the two monthly FRED series, with gray bands marking NBER recessions. \textbf{C}: the coupling plan as a transport matrix. (i)~the mediated plan $T_{ij}=T_jT_i^{\top}$ on a logarithmic scale in units of the uniform coupling $U$, spanning $10^{-2}U$ to $10^{2}U$, collapses to a sparse matrix in which a few support pairs carry mass two decades above uniform (normalized mutual information $0.24\pm0.01$ against a random-walk null of $0.064$). Bright cells locate the correspondence while the depth of the dark field measures its exclusivity, because the doubly stochastic marginals force every excess above uniform to be paid for by depletion across the same row and column. (ii)~the same coupling on rolling 10-yr windows, which stays above the null in every regime, with NBER recessions again in gray. }
\label{fig:phillips}
\end{figure*}

The joint $(\mathrm{UNRATE},\,\pi_{\mathrm{YoY}})$ Phillips attractor~\cite{phillips1958relation}, computed from monthly FRED data (Fig.~\ref{fig:phillips}B), supplies the first real-data test of the PSoPS basis as a structural detector. We treat the joint cloud directly as a metric-measure space and ask whether it carries a persistent loop (Vietoris--Rips $H_1$, computed via \texttt{ripser}~\cite{bauer2021ripser} and stability of persistence diagrams under perturbation guaranteed by Cohen-Steiner et al.~\cite{cohen2007stability}). Persistent homology has previously been applied to financial time series to detect the topological signature of market crashes~\cite{gidea2018tda}.

Here we ask, do the unemployment and inflation attractors share geometric structure? We build each series' attractor (Takens $\mathrm{dim}=8$, $\tau=4$, per-series diffusion map with $n_s=40$ farthest-point supports in accordance with the construction guidelines stated above), take the pairwise Gromov--Wasserstein barycenter of the two, and read off the mediated transport plan $T_{ij}=T_jT_i^{\top}$ (Fig.~\ref{fig:phillips}C(i)). If the two series predict one another, this plan should collapse toward a sparse, near-permutation matrix as each support in one maps to a support or combination of supports in the other transport matrix cleanly. This is precisely what we find with normalized mutual information of $0.24\pm0.01$ over 12 entropic-GW replicate seeds, against $0.064$ for an unrelated random-walk null run through the identical pipeline, whose spread over the same $12$ seeds is below $10^{-3}$. The directed coupling $\mathcal{S}_{i\to j}$ (Eq.~\ref{eq:sensitivity}) is correspondingly concentrated. The participation ratio of the coupling tensor, $\mathrm{PR}(\mathcal{S}) = (\sum_{kl}\mathcal{S}^{kl})^2/\sum_{kl}(\mathcal{S}^{kl})^2$, counts how many of the $25$ mode pairs carry appreciable weight. It is distinct from the spectral participation ratio of Table~\ref{tab:law-geometry}, which is computed on diffusion eigenvalues rather than on coupling entries. Here it is $6.8\pm0.7$ versus $11.2\pm0.6$ for the null while the coupling is weakly asymmetric ($\lVert\mathcal{S}_{i\to j}-\mathcal{S}_{j\to i}^{\top}\rVert/\lVert\mathcal{S}_{i\to j}\rVert=0.13$), which is the geometric signature of a lead--lag relationship. The asymmetry has a direction: unemployment explains inflation better than the reverse, $\lVert\mathcal{S}_{\mathrm{UNRATE}\to\pi}\rVert / \lVert\mathcal{S}_{\pi\to\mathrm{UNRATE}}\rVert = 1.12$, holding on all $12$ barycenter seeds (paired $t$-test $p<10^{-11}$). Unemployment leads and inflation lags, recovered here from the geometry alone.
We also find strong loop structure in each geometric object: unemployment attractor ($\beta_1=43$), the inflation attractor ($\beta_1=54$), and their joint embedding ($\beta_1=39$, Fig.~\ref{fig:phillips}A). These counts are much larger than the $\beta_1$ reported for Phillips in Tables~\ref{tab:deff} and \ref{tab:law-geometry} because they count loops in a different object. Here the input is the three-dimensional diffusion-map embedding of the delay attractor, built on the full monthly record of $936$ observations, which resolves every business cycle in that record as its own cycle along with the smaller excursions between them. There the input is the raw two-dimensional (unemployment, inflation) cloud after cyclical detrending, on which only the handful of loops that are subsampled are counted. Both are reported because they answer different questions: whether the coupled dynamics are looped at all, and how many loops are robust features of the law itself. 
  This persistent homology is characteristic of business-cycle hysteresis carried by the coupled geometry and it persists across regimes (Fig.~\ref{fig:phillips}C(ii)).

\subsection{The universal Phase Space of Phase Spaces}
\label{subsec:universal-aoa}\label{sec:structure}

This section establishes the central empirical result of the paper: the dynamics of the macroeconomy are embedded in a single connected geometric object of effective dimension $7.3\pm0.2$, and its named concepts are enriched regions of that object rather than separable clusters. We construct a corpus of macroeconomic data from 14 providers using four deterministic rules to subsample to a core of data small enough to build the PSoPS on. First, we restrict data to economic sources and hold back health, weather and river-flow data as out-of-domain controls, revisited at the close of this section. Second, a series must carry enough observations for a delay embedding to sample its attractor, and its cost matrix, the unit-max diffusion distance matrix of \S\ref{subsec:dmap}, must be non-degenerate. Third, only macroeconomic reporting frequencies are kept, and where the same underlying series is published at several frequencies we retain the best-resolved one. Fourth, no single macro concept (rates, employment, output and so on) may exceed twelve percent of the candidate pool, which stops one heavily-published family from defining the geometry. The cap is applied by a seeded uniform draw within any over-represented concept, so the rule is reproducible. It binds on exactly two groups, trimming interest rates from $5{,}922$ series and the residual ECB group from $5{,}411$ down to the cap of $2{,}744$ each, and because trimming shrinks the pool the realized share of the largest concept in the final corpus is sixteen percent. Twelve percent is a choice rather than a derived quantity. It is the loosest cap under which no concept exceeds twice the share of the next largest and the measured dimension does not turn on it (Fig.~\ref{fig:aoa-robustness}). Applying these at an observation threshold of $60$ yields $17{,}025$ series, admitting the annual country-level macro (World Bank output and labor, Penn World Table productivity and capital). Eurostat, which is the large majority of the raw cache and is granular regional and sectoral rather than macroeconomic, is excluded from the core and re-attached afterwards by Nystr\"om extension (\S\ref{subsec:extension}). The restriction and the re-attachment are two steps, not one: the geometry is fitted without Eurostat and then asked to place it. 

Applied to the global macroeconomy, the pipeline ingests $281{,}536$ series. Of these, $280{,}149$ are economic, drawn from 14 providers and spanning $222$ national economies together with the World Bank regional and income aggregates, the euro area and the European Union. The remaining $1{,}387$ are the health, weather and river-gauge streams held back as out-of-domain controls. 
 Transport plans are solved for the $17{,}025$ series that pass the selection rules above, and it is these that collapse onto a single low-dimensional submanifold. We test the basis on three textbook relations whose qualitative geometry is already understood, so a faithful basis has something concrete to reproduce. The Phillips curve (unemployment against inflation) should trace a hysteresis loop as the economy cycles through expansion and contraction. Okun's law (output against the change in unemployment) carries the same boom--bust loop alongside direct exposure to growth factors and the monetary transmission chain. Finally the Solow model (decomposition of GDP) is a smooth, strongly curved monotone. If the PSoPS is a genuine basis for the macroeconomy it will recover a loop where the economics says there is a loop and curvature where the economics says there is curvature. This is a direct geometric interpretation of economic law and the PSoPS merely expresses that geometry from the data.

In order to deal with this intractably large set of pairwise transport plans (grows like $N^2$), we compute per-series GW for every series, with the Hellinger-dMap embedding built on 800 FPS landmark plans and Nystr\"om-extended to the rest onto a low-dimensional submanifold of the Birkhoff polytope. This object is our geometric basis, of effective dimension $d_{\mathrm{eff}} = 7.3 \pm 0.2$ determined by the bandwidth-free Coifman $\varepsilon$-scaling dimension~\cite{coifman2006diffusion} evaluated on the whole corpus, with the point estimate on all $17{,}025$ series and the uncertainty from eight bootstrap draws of $6{,}400$ series taken without replacement. 
 This dimensionality is robust to corpus composition. Draws taking equal numbers of time series from every concept return $7.13\pm0.00$ against $7.27\pm0.05$ for unrestricted draws of the same size. Adding new data densifies the manifold rather than moving its axes.

The dimension also resolves structure within the corpus. Holding the sample size fixed at $2000$ series so that the comparison is not confounded by the estimator's own size dependence, the monetary and rate complex returns $8.46\pm0.06$ (rates alone) and $8.43\pm0.05$ (rates with credit), exchange rates $6.63\pm0.01$, and real activity, meaning output, labor, inflation, unemployment, capital and productivity together, only $5.57\pm0.08$. Against a draw-to-draw scatter of $0.05$ these are not close. The price-of-money side of the economy carries roughly three more independent dynamical modes than the real side.

\begin{figure*}[!tp]
\centering
\includegraphics[width=\textwidth]{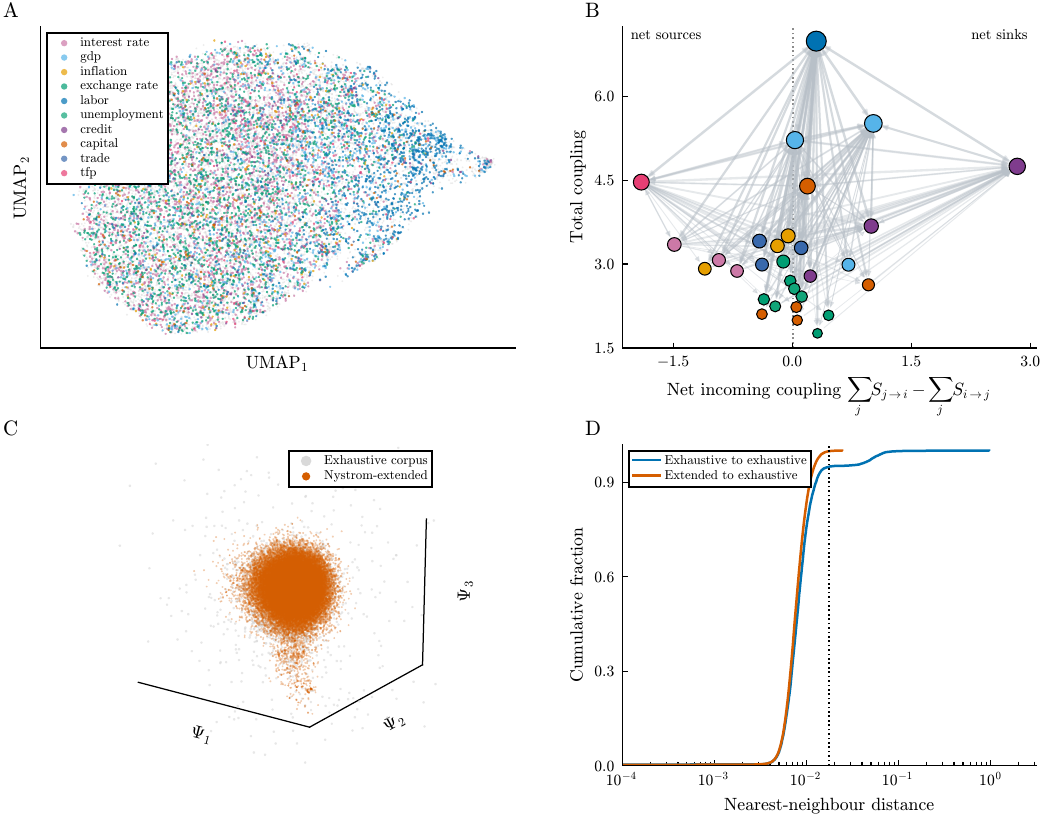}
\caption{\textbf{The macroeconomy is one geometric object.} \textbf{A}: UMAP embedding of the per-series PSoPS coordinates of the exhaustive corpus (all $17{,}025$ series, ten retained coordinates each), colored by economic concept with the residual groups in gray. Named concepts occupy overlapping, enriched regions of a single connected sheet rather than separable clusters. \textbf{B}: the directed coupling network of the $30$ conceptual barycenters, colored by concept as in A and sized by total coupling. Horizontal position is the net incoming coupling, so net sources sit left of the dotted zero line and net sinks right. Arrows mark the top $30\%$ of directed couplings $S_{i\to j}$, and $168$ of the $263$ drawn arrows flow from the source side toward the sink side. Against a null that swaps $S_{i\to j}$ and $S_{j\to i}$ at random for every pair and recomputes the layout, that fraction ($0.64$) exceeds every one of $2000$ null draws (null $0.53\pm0.01$). \textbf{C}: the Nystr\"om-extended corpus ($381{,}677$ phases spaces due to multiple time delays, orange) lands inside the support of the exhaustive corpus ($17{,}025$ series, gray) in the same coordinates. \textbf{D}: nearest-neighbor distance CDFs. The extended-to-exhaustive curve saturates at the $95$th percentile of the exhaustive corpus's own nearest-neighbor spacing (dotted), with $99.6\%$ of extended series within the support. }
\label{fig:universal-aoa}
\end{figure*}

\begin{figure*}[!tp]
\centering
\includegraphics[width=\textwidth]{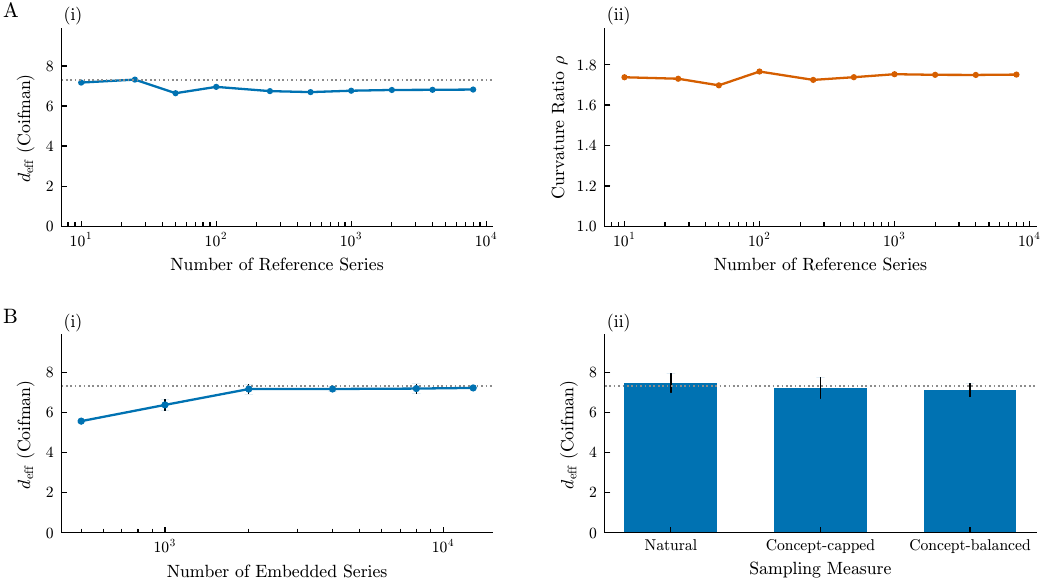}
\caption{\textbf{The universal PSoPS is stable.} The dotted line in every dimension panel marks the headline $d_{\mathrm{eff}}=7.3$. \textbf{A}: Fr\'echet-mean scaling. (i)~the Coifman $d_{\mathrm{eff}}$ of a fixed $1500$-series evaluation panel as the number of reference series building the barycenter grows from $10$ to $8000$, and (ii)~the curvature ratio $\rho$ of the same panel, the ratio of geodesic to straight-line distance. Both are flat, so the reference is in its asymptotic regime. The level in (i) sits below the $7.3$ line because the estimator is evaluated on the fixed panel and reads low at small $n$ (\S\ref{subsec:universal-aoa}). \textbf{B}: dimension robustness on the full Hellinger coordinates. (i)~$d_{\mathrm{eff}}$ against the number of embedded series, with error bars the standard deviation over independent draws, converging onto the headline line from $n=2000$ up. (ii)~$d_{\mathrm{eff}}$ at matched $n=3000$ under three sampling measures, the natural corpus draw, the concept-capped draw of \S\ref{subsec:universal-aoa}, and a concept-balanced draw. Composition moves the estimate by less than the draw scatter.}
\label{fig:aoa-robustness}
\end{figure*}

We read this alongside the coupling network below (Fig.~\ref{fig:universal-aoa}B), in which the interest-rate barycenters are the most consistently directional sources. Two unrelated measurements, one of coupling direction and one of dimension, identify the same part of the macroeconomy as the richest. A basis that returned the same dimension for every sub-domain would be one that had failed to resolve this. 
 

We visualize the PSoPS with a UMAP~\cite{mcinnes2018umap} of the per-series PSoPS coordinates (Fig.~\ref{fig:universal-aoa}A), showing a single connected sheet (one connected component at the percolation scale) on which named economic concepts (interest rates, GDP, inflation, exchange rates, labor, credit, $\ldots$) occupy overlapping but enriched regions. This is the geometric face of the central thesis. The high degree of coupling across the macroeconomy results in a single strongly-coupled low-dimensional dynamical object, not a collection of independent factors. This single object is what makes the tangle of macroeconomic series tractable.

The coupling structure is recovered at the concept level by the network of conceptual barycenters $\{B^{*}_i\}$. Thirty barycenters, one for each of ten named indicators (capital, credit, exchange rate, GDP, inflation, interest rate, labor, TFP, trade and unemployment) at each reporting timescale it is published at, from annual down to daily for the market-linked series, together with the directed coupling tensor $S_{ij}$ on transport-plan entropy, define a coupling network laid out by direction in Fig.~\ref{fig:universal-aoa}B, where a node's horizontal position is its net incoming coupling, so net sources sit on one side of zero and net sinks on the other and the layout itself carries the finding. At the concept level the coupling matrix is estimated spectrally, from the eigenvalue-weighted association between the diffusion-map axes of barycenter blocks, probed at $\varepsilon=0.1$ with an entropy gate that skips near-uniform plans. This is a coarse-grained implementation of the sensitivity tensor of Eq.~\ref{eq:sensitivity}, which is evaluated exactly at the law level in \S\ref{subsec:phillips-loop}. Arrows are drawn between pairs whose directed coupling $S_{ij}$ falls in the top 30\% of all pairs, and $168$ of the $263$ drawn arrows flow from the source side toward the sink side. That threshold is a legibility choice for the drawing and nothing below is read from it, since the quantities we interpret are sums over all directed pairs and do not depend on which edges are shown. Two such quantities organize the network. Total coupling is carried chiefly by the real aggregates, with labor and GDP the most connected nodes, which is what it means for them to be the most explained quantities in the economy. The direction of coupling tells the sharper story. Ranking nodes by net outgoing coupling, how much a node explains the others beyond what they explain of it, the top five sources are productivity, the interest rate at each of its three timescales, and monthly inflation. The rate nodes rank $2$, $4$ and $5$ of $30$, a concentration whose chance probability is $2.5\times10^{-3}$. The deepest sinks are credit, output and capital. We note that this ledger is zero-sum by construction, since every unit of outgoing coupling at one node is incoming coupling at another, so source and sink describe the asymmetry of explanation between fluctuation dynamics, the lead and the lag, and carry no statement about growth. The price of money and technology drive, and credit and output are driven, which is the credit channel of monetary transmission recovered as a direction rather than a degree~\cite{bernanke1995inside}. In this sense the basis does not merely compress the macroeconomy, it orients it. Each concept receives a place in the economy's information flow, and the textbook transmission chain appears as the geometry's arrow of explanation rather than as an estimated model. The direction is not a bookkeeping artifact of the layout. Under a null that swaps $S_{i\to j}$ and $S_{j\to i}$ at random for every pair and recomputes the layout, the observed fraction of strong couplings flowing from the source side toward the sink side, $0.64$, exceeds every one of $2000$ null draws (null $0.53\pm0.01$). The dimensional result above points the same way from an unrelated measurement, since the monetary sub-corpus is also the highest-dimensional one, so the channel that transmission theory places at the origin is both the most directional and the richest part of the geometry. However, we note that this reading is a property of this indicator set, not a proof that no other organization exists. 
Constructing this coarse-graining without the indicator prior does not recover the named concepts as distinct groups (adjusted Rand index $\approx0.009$ against the indicator labels, quantified below), because the macroeconomic manifold is one connected, overlapping object rather than a set of separable clusters. The conceptual barycenters are therefore a deliberate human coarse-graining of a continuum, not a discovered partition, and we retain the network for the transmission reading with that caveat.

The PSoPS above is built on a single universal Fr\'echet-mean barycenter (a true GW barycenter, not a medoid), to which every series is transported. We tested whether per-concept (``many'') barycenters would tell a different story. They do not, and the comparison is itself a result. (i) Timescale-invariance: barycenters built separately on annual, quarterly, and monthly series are nearly coincident. The silhouette score~\cite{rousseeuw1987silhouettes} of the timescale grouping is $-0.015$, a dimensionless index on $[-1,1]$ comparing within-group to nearest-other-group distance, and a silhouette at or below zero means points sit no closer to their own group than to a neighboring one, so the three timescales do not form separable clusters at the macroeconomic reporting timescale, and resampling frequency does not change the geometry. That would not hold for micro data, where intraday FX prices live on different dynamics, but this macroeconomic universal barycenter loses nothing by pooling timescales. (ii) Concepts are not discovered clusters: a purely data-driven coarse-graining of the PSoPS coordinates (10-way clustering by dynamical similarity alone) does not recover the named indicators, adjusted Rand index $\approx0.009$ against concept labels. (iii) Per-concept sub-manifolds differ, but by less than concept size alone would suggest. Concepts here range from $190$ to $2744$ series, and the estimator is size-dependent across exactly that range. Drawing every concept at a common $n$ instead, the local effective dimension runs from $3.57$ for capital to $5.79$ for the World Bank residual group, with labor $5.66$, productivity $5.59$, interest rates $5.31$, GDP $4.33$ and inflation $3.94$. No concept is degenerate, but none spans the full geometry either. 
 The only robust structure the many-barycenter network adds beyond the single object is the directed source and sink layout of the coupling network (Fig.~\ref{fig:universal-aoa}B). The single universal barycenter is therefore the canonical object for the global macroeconomy. Source: \texttt{pipeline/coordinates.py}, \texttt{pipeline/barycenter\_plans.py}.

The headline estimate of $7.3$ should be read as a floor. Calibrating the estimator against $d$-spheres of known dimension embedded in the same $1600$-dimensional ambient space and evaluated at the same sample sizes shows that the corpus reads what a nine-dimensional object reads, and inverting the calibration curve implies a true intrinsic dimension near $9.5$ (SI). We quote the measured $7.3$ because the calibration uses uniformly sampled spheres and the corpus is neither uniform nor spherical. The stricter $n\ge200$ gate points the same way, returning $8.4\pm0.3$ on its $11{,}928$ longer series, which is the direction the floor argument predicts. 

This floor makes the comparison with the dynamic factor literature quantitative rather than rhetorical. Phase-randomized surrogates that preserve the entire linear covariance structure destroy $24$--$44\%$ of the measured dimension (\S\ref{subsec:nonlinearity}), so a second-order method can see at most $56$--$76\%$ of the geometry however many factors it retains. Applied to a $9.5$-dimensional object, that predicts a linear method should recover between five and seven factors, which is what the factor-model literature reports. The DFM count is therefore not a rival estimate of the same quantity but the shadow a linear method casts on this one, and the discrepancy is the part of the macroeconomy that is not expressible in a linear basis.

Convergence is reached well before the full corpus. The estimate climbs by ${\sim}0.8$ per doubling of $n$ up to $n\approx1600$ and by ${\sim}0.1$ thereafter, with the increments changing sign above $n=3200$, so the remaining movement is sampling scatter rather than drift. The Fr\'echet-mean reference is stable as the corpus grows from $\sim$10 to $\sim$8000 series, in both its measured dimension and its curvature ratio, the ratio of geodesic to straight-line distance (Fig.~\ref{fig:aoa-robustness}A(i)--(ii)). The barycenter and the resulting $d_{\mathrm{eff}}$ plateau well before the full corpus, so the construction is in its asymptotic regime and the over-represented rate data densifies rather than reshapes it. And the dimension is robust across corpus size and sampling measure (Fig.~\ref{fig:aoa-robustness}B), with the draw-to-draw scatter small against the estimate itself. Barycenters built from two independent halves of the corpus are nearly indistinguishable after optimal alignment, with a Procrustes disparity of $0.09$ on a $[0,1]$ scale and kernel eigenvalue spectra agreeing to within $4\%$ over the leading ten modes. 

The corpus construction holds back $156$ out-of-domain controls ($106$ WHO health, $30$ river-flow hydrology, $20$ NOAA weather), which enter no macroeconomic result. They also delimit what the basis is. Under matched solver settings the controls are not rejected by the construction: their transport plans carry structure comparable to macroeconomic plans at every common $\varepsilon$ tested (structure ratios of $1.1$ to $1.5$ with AUC at or below $0.59$), and placed by the extension machinery of \S\ref{subsec:extension} they land within the manifold support at a comparable rate ($98.7\%$, against $99.6\%$ for the held-out Eurostat reference).  The PSoPS is a representation of attractor geometry, not of domain membership. Any well-sampled dynamics resolves in it, and what is economic about the object built here is the corpus it is built from, not a filter the construction applies.

\subsection{Extending the basis to the full corpus}
\label{subsec:extension}

The manifold is fitted on the $17{,}025$ series that pass the selection rules, but the pipeline ingests $281{,}536$. A basis is only worth the name if it can place data it was not fitted on, so we run the extension the corpus rules were designed to permit. Every held-out series still needs its own entropic-GW solve against the frozen universal barycenter. What Nystr\"om avoids is the dense $N\times N$ eigendecomposition, which at this scale is the intractable step. Of the $399{,}260$ held-out Eurostat attractors at macroeconomic reporting frequency with $n\ge60$ observations, $381{,}677$ yield a non-degenerate cost matrix and are placed in the PSoPS coordinates of the exhaustive corpus. The extension itself is validated wherever a dense computation is available for comparison. On the macroeconomic monetary domain the FPS-landmark Nystr\"om recovers the leading PSoPS modes of the full dense Hellinger-DMAP with per-mode correlation $\rho_1 = 0.97$ and $\rho_2 = 0.93$ (validation run at $\varepsilon=0.05$, the graded-similarity regime), and on the controlled chaotic-systems corpus, where ground truth is known, the agreement is $\rho = 0.88$ averaged over all retained modes. 

They land on the object: $99.6\%$ of extended series fall within the support of the exhaustive corpus, and the median nearest-neighbor distance from an extended series to an exhaustive one ($0.0076$) sits just under the typical nearest-neighbor distance inside the exhaustive corpus itself ($0.0080$, Fig.~\ref{fig:universal-aoa}D). The effective dimension is preserved: measured on the same ten PSoPS coordinates, the extended cloud gives $5.42$ against $5.10$ for the exhaustive corpus. These two numbers are lower than the headline $d_{\mathrm{eff}} = 7.3 \pm 0.2$ because they are measured on the ten retained coordinates rather than on the full $1600$-dimensional transport plans. The comparison that matters here is between them, not against the headline.

The extended series also spread as the corpus does (Fig.~\ref{fig:universal-aoa}C). The interquartile span of the extended cloud matches the exhaustive corpus axis by axis across the ten retained coordinates, with the largest deviation at twelve percent. This deserves a check, because a Gaussian-kernel Nystr\"om extension shrinks a point toward the origin when that point is far from every landmark: the kernel row goes near-uniform and the projection loses contrast whatever the underlying geometry. That mechanism has nothing to act on here. The held-out series sit at a median distance of $0.982$ from the nearest landmark against $0.983$ for the exhaustive corpus, so the two populations are covered equally well. Regional and sectoral Eurostat data neither collapses to a core of the macroeconomic geometry nor scatters off it. It fills the same object.

This is the sense in which the paper's claim over the full corpus should be read. The geometry is measured on $17{,}025$ series and the remaining $381{,}677$ are placed on it, at the same effective dimension and inside the same support, without re-fitting anything. The extension adds density, not axes, which is what a basis is supposed to do. 

\begin{table*}[tp]
\centering
\caption{\textbf{Intrinsic dimensionality of the universal manifold and the canonical laws.} The Coifman $\varepsilon$-scaling dimension~\cite{coifman2006diffusion} is the estimator of record, continuous and bandwidth-free. The Grassberger--Procaccia correlation dimension $D_2$~\cite{grassberger1983characterization} is an independent estimator reported only where a scaling region exists, which excludes Okun and the universal manifold (see text). The diffusion-mode count is the L-method knee~\cite{salvador2004determining} on the eigenvalue spectrum, an embedding count rather than a dimension, and $\beta_1$ is the bootstrapped count of significant persistent loops, on the cyclical-detrended plane for the laws and on the diffusion coordinates for the universal manifold. Law windows are Phillips 1957--2025 monthly, Okun 1948--2025 quarterly, Solow 1947--2025 quarterly. }
\label{tab:deff}
\begin{ruledtabular}
\begin{tabular}{lcccc}
Object & Coifman $d_{\mathrm{eff}}$ & $D_2$ & Diffusion modes & $\beta_1$ \\
\hline
Universal PSoPS & $7.3\pm0.2$ & --- & $3.9\pm1.3$ & $4.8\pm1.3$ \\
Phillips & $1.87\pm0.02$ & $1.59\pm0.02$ & $5.9\pm1.9$ & $7.2\pm1.0$ \\
Okun & $2.40\pm0.02$ & --- & $10.0\pm2.2$ & $3.7\pm1.0$ \\
Solow & $1.92\pm0.02$ & $2.38\pm0.07$ & $7.5\pm1.7$ & $3.0\pm0.8$ \\
\end{tabular}
\end{ruledtabular}
\end{table*}

\begin{table}[tp]
\centering
\caption{\textbf{Per-law geometry and topology.} Participation ratio PR (effective number of diffusion modes), Hellinger/Frobenius energy ratio H/F, and curvature ratio $\rho$ (geodesic/Euclidean) on each law's barycenter, built from five member series per law. The persistent first Betti number $\beta_1$ is the \emph{bootstrapped} significant-loop count on the full cyclical-detrended attractor (Fig.~\ref{fig:law-dim}B). All three laws loop as expected while PCA on the same panels resolves none. }
\label{tab:law-geometry}
\begin{ruledtabular}
\begin{tabular}{lcccc}
Law & PR & H/F & $\rho$ & $\beta_1$ \\
\hline
Phillips & $3.84$ & $0.78$ & $1.17$ & $7.2\pm1.0$ \\
Okun & $3.97$ & $0.60$ & $1.07$ & $3.7\pm1.0$ \\
Solow & $3.39$ & $0.76$ & $1.65$ & $3.0\pm0.8$ \\
\end{tabular}
\end{ruledtabular}
\end{table}

\subsection{Macroeconomic Dynamics are Spanned by a Nonlinear Basis}
\label{subsec:nonlinearity}

\begin{figure*}[!tp]
\centering
\includegraphics[width=\textwidth]{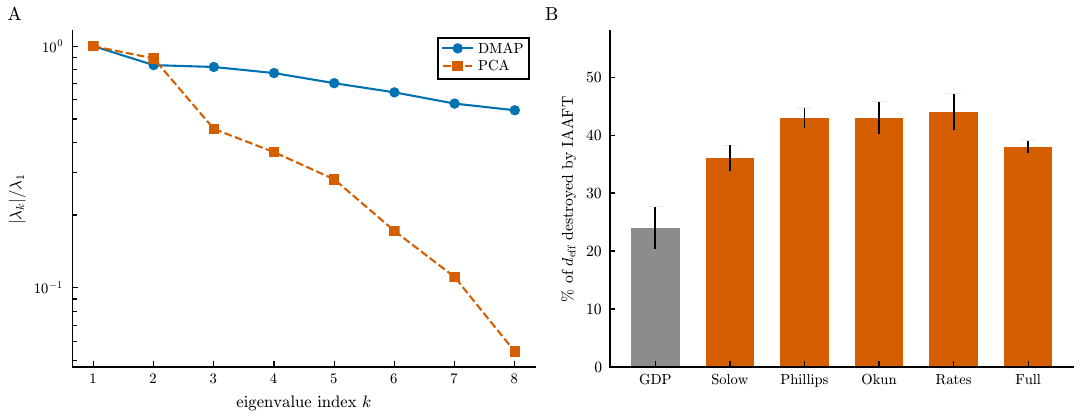}
\caption{\textbf{The PSoPS basis is nonlinear.} \textbf{A}: normalized eigenvalue spectrum of the per-series diffusion map (blue) versus PCA (orange) on a Phillips member series (core CPI). The nonlinear basis carries information far down the tail while PCA decays by more than an order of magnitude. \textbf{B}: IAAFT phase-randomized surrogate destruction of $d_{\mathrm{eff}}$, with error bars the standard error of the mean over 10 surrogate realizations. Across the six test cases the linear-preserving surrogate destroys $24$--$44\%$ of the dimensionality. The five colored cases fall outside the one-sided 99\% surrogate band, while GDP ($z=2.1$) sits just inside.
}
\label{fig:nonlinearity}
\end{figure*}

The dimensional and topological compactness reported above is a property of the nonlinear basis. The same timeseries projected onto a linear PCA basis are diffuse and high-dimensional. Two diagnostics make this concrete (Fig.~\ref{fig:nonlinearity}). First, on a Phillips member series (core CPI), the per-series diffusion-map spectrum keeps a heavy tail across the retained modes while the PCA spectrum of the same delay cloud falls by more than an order of magnitude within eight modes. The nonlinear basis sees the slow modes that linear factors do not. 

Second, we test whether the recovered geometry is itself nonlinear or only a linearly-deformed Gaussian by phase-randomizing the input series with iterative amplitude-adjusted Fourier transform (IAAFT) surrogates and re-running the pipeline. The IAAFT builds surrogate series that keep the original power spectrum and value distribution exactly but scramble the Fourier phases~\cite{schreiber1996improved}. Anything that remains in the surrogate is explainable by linear (second-order) statistics alone, so whatever the surrogate destroys was genuinely nonlinear. We run six test cases: the three canonical laws, two further macroeconomic relations, and the full corpus. The IAAFT pipeline destroys $24$--$44\%$ of the effective dimensionality (Phillips $-43\%$, Okun $-43\%$, Solow $-36\%$, rates $-44\%$, GDP $-24\%$, full corpus $-38\%$), and the observed $d_{\mathrm{eff}}$ falls outside the one-sided 99\% surrogate band in five of the six cases, all but GDP, whose $z$-score of $2.1$ sits just inside. The three canonical laws span the narrower range $36$--$43\%$. Linear surrogate data cannot reproduce the PSoPS geometry. Source: \texttt{scripts/\allowbreak fig6\_nonlinearity.py} (panel~A: live Takens embed + DMAP vs.\ PCA on FRED \texttt{CPILFESL} log-returns, panel~B: IAAFT \%-destruction loaded live from atlas \texttt{iaaft\_timeseries\_*.json}).

\subsection{Macroeconomic laws as per-law attractors}
\label{subsec:law-attractors}

The participation ratio $\mathrm{PR}=(\sum_k\lambda_k)^2/\sum_k\lambda_k^2$ counts how many diffusion modes carry real weight. A flat one-mode object has $\mathrm{PR}\approx1$ while a multi-dimensional one has $\mathrm{PR}\gg1$. The curvature ratio $\rho$ divides geodesic distance along the manifold by straight-line Euclidean distance, so $\rho=1$ is flat and $\rho>1$ flags a curved sheet that a linear basis would distort. The first Betti number $\beta_1$ counts persistent loops. The Hellinger-to-Frobenius energy ratio H/F reports the realized Hellinger amplification as a fraction of the theoretical $n_s/2$ bound of \S\ref{sec:gw}, so the laws realize $60$--$78\%$ of the maximum resolution gain.

\begin{figure*}[!tp]
\centering
\includegraphics[width=\textwidth]{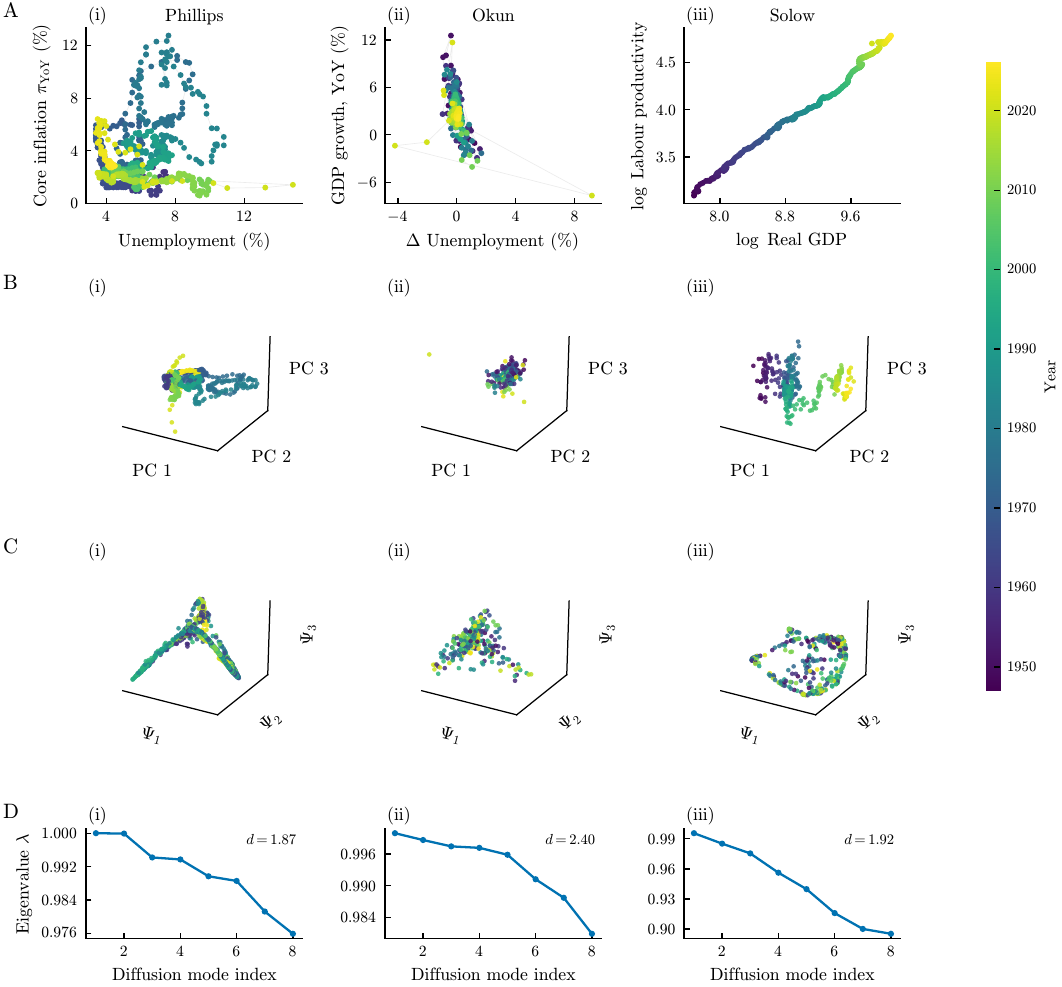}
\caption{The canonical laws are curved, looped attractors, and the curvature is invisible to a linear reduction. Columns are Phillips, Okun and Solow with color corresponding to year, on one scale across all panels. Row A: the raw joint plane of each law, threaded by a faint gray path in time order. Row B: the joint delay embedding of the same two series reduced to three dimensions by PCA. Row C: the same delay embedding reduced by the diffusion map. Rows B and C differ only in the reduction, so the comparison isolates linearity. The linear projection collapses each law toward a diffuse cloud, while the nonlinear one resolves the loop. The curvature ratio, participation ratio and bootstrapped loop count for each law are tabulated in Table~\ref{tab:law-geometry}. Row D: the Hellinger-diffusion eigenvalue spectrum of each law, annotated with its Coifman dimension. The spectra are where the loop shows up independently of any homology computation: a closed curve has degenerate diffusion eigenfunctions $\cos k\theta$ and $\sin k\theta$, so Phillips arrives in near-equal pairs at modes 1--2, 3--4 and 5--6, Okun in one pair, and Solow, the curved monotone rather than a loop, in none. Source: \texttt{scripts/fig7\_law\_attractors.py}, \texttt{validation/law\_dimensionality.py}.}
\label{fig:law-dim}
\end{figure*}

The canonical macroeconomic laws~\cite{phillips1958relation,okun1962potential,solow1956contribution} are themselves geometric objects when read from raw FRED data (Fig.~\ref{fig:law-dim}A). The Phillips curve lives on the $(\mathrm{UNRATE},\,\pi^{\mathrm{core}}_{\mathrm{YoY}})$ joint plane (monthly, 1957--2025, $n=812$), Solow on the $(\log\mathrm{GDPC1},\,\log\mathrm{OPHNFB})$ plane (quarterly, 1947--2025, $n=312$), and Okun on the $(\Delta\mathrm{UNRATE},\,\mathrm{GDP}_{\mathrm{YoY}})$ plane (quarterly, 1948--2025, $n=308$). Applied to the law-scoped corpora (five member series per law) the PSoPS construction returns curvature on all three ($\rho_{\mathrm{geo/eucl}} > 1$). Solow has the most strongly curved manifold ($\rho = 1.65$, the productivity--output monotone), Phillips is intermediate ($\rho = 1.17$) and Okun is near-flat ($\rho = 1.07$) but with the highest participation ratio ($\mathrm{PR} = 3.97$, Table~\ref{tab:law-geometry}). All three laws are looped, which is the geometric form of hysteresis: a relation whose path out of a downturn differs from its path in, so that the joint cloud cannot be the graph of any single-valued function~\cite{blanchard1986hysteresis}. The loops agree with the business-cycle hypothesis, with $\beta_1 = 7\pm1$ (Phillips), $3\pm1$ (Solow), $4\pm1$ (Okun) over 40 subsamples. These business-cycle loops are visible to the nonlinear basis and absent from a linear projection of the same panels, as established in \S\ref{subsec:nonlinearity}. 

Applying the same bandwidth-free dimensionality analysis as the universal manifold (Coifman $\varepsilon$-scaling, with persistent-homology loop counts on the detrended cyclical attractors), the three laws are confirmed to be genuinely low-dimensional (Fig.~\ref{fig:law-dim}). Detrending is essential for Solow. The raw $(\log\mathrm{GDPC1},\log\mathrm{OPHNFB})$ levels co-trend into a spurious one-dimensional line ($d\approx1.1$), and only the cyclical component reveals the true $\sim$2-D dynamics. 

The diffusion-mode counts of Table~\ref{tab:deff} are the corresponding finite-size diagnostic. The L-method elbow on a 60-point DMAP eigenvalue spectrum quantizes to integers, so the law-scoped count is reported only as a lower bound, $d_{\mathrm{eff}}^{\mathrm{LM}} \geq 2$ at $N \sim 60$. We quote the continuous Coifman $\varepsilon$-scaling estimate for structural claims and report the L-method count beside it, for the reasons set out in \S\ref{sec:gw} and the SI. The L-method count exceeds $d_{\mathrm{eff}}$ on the laws because a closed loop yields degenerate harmonic pairs $\cos k\theta$, $\sin k\theta$, so an embedding count double-books each loop dimension. Where the correlation dimension $D_2$ is defined it brackets the Coifman value, placing Phillips and Solow near two dimensions. It is undefined for Okun, not for want of points but because its differenced, noise-dominated plane never opens a scaling region under the $R^2>0.95$ gate, while Solow at nearly the same sample size traces a smooth monotone curve and does. It is likewise undefined for the universal manifold, where the $1600$-dimensional unit-norm plans concentrate distances and the local slope never forms a plateau. The Coifman estimator reads the steepest point of the curve rather than requiring a flat one and consequently can be computed in all cases. On the universal manifold the count runs the other way. The ${\sim}4$ variance-dominant diffusion modes of Table~\ref{tab:deff} are an embedding count of the variance-dominant directions, and the gap to $d_{\mathrm{eff}}=7.3$ is genuine lower-variance sub-structure rather than a disagreement between the two quantities.

We have shown two things. Firstly, the whole macroeconomy (up to our sampling) is described by a single low-dimensional geometric object. The $17{,}025$ embedded series collapse onto a single connected object of effective dimension $d_{\mathrm{eff}}=7.3\pm0.2$, and this is robust to composition, with concept-balanced draws returning $7.13$ against $7.27$ for unrestricted draws of the same size. Secondly, each named law lives as a $\sim$2-dimensional looped sub-attractor inside it (Phillips, Okun, Solow all $d\approx2$, carrying multiple business-cycle loops). We conclude that canonical laws are not separate models bolted together but low-dimensional, cyclical neighborhoods of one connected geometry. The macroeconomy is organized by a handful of shared modes and the named laws are its projections.

\subsection{Velocity and regime detection}\label{subsec:velocity}\label{sec:prediction}\label{sec:velocity}

Takens' theorem assumes ergodicity, but macroeconomic systems evolve. Tracking the velocity of the per-series transport plans $\dot T_i$ therefore turns the basis into a regime-change detector: a sharp step in $\dot T_i(t)$ marks a date at which the dynamical basis itself reorganized. Two scales emerge.

\subsubsection{Acceleration in transport plan entries marks canonical regime transitions}
\label{subsubsec:velocity-positive}

The transport plan $T_i(t)$ is recomputed on a rolling window, so its rate of change measures how fast the relational geometry of a series is deforming. We call this the transport-plan velocity, $v(t)=\lVert T_t-T_{t-\Delta}\rVert_F$, and a step in $v(t)$ marks a date at which the basis itself reorganized rather than one at which a fitted coefficient moved. The control is a permutation null: the time index is shuffled $1000$ times to destroy temporal order while preserving the marginal distribution of plan entries, so any step remaining after the shuffle cannot be an artifact of the plan distribution alone. On the universal PSoPS this null, with Benjamini--Hochberg control of the false discovery rate at level $q=0.05$, which caps the expected fraction of spurious detections across the indicators \cite{benjamini1995controlling}, certifies a velocity step in 6 of 7 macroeconomic indicators (GDP, Unemployment, Inflation, Interest, Full, Solow, Phillips) between 1970 and 2022, peak $\sigma = 4.9$ above the null. The detected dates are not arbitrary. They align with the textbook regime anchors of postwar monetary history. These are (1) the 1978 inflation run-up that the Volcker disinflation answered (Solow, $\sigma=4.9$, detected 1978-05), (2) the Draghi 2012 ``whatever it takes'' pledge (Unemployment, Phillips, Full with $\sigma \in [4.5, 4.8]$), (3) COVID 2020 (Inflation, $\sigma=3.9$) and (4) the 2006 turn into the GFC (GDP, $\sigma=3.4$). The Interest indicator at 2014 ($\sigma=2.3$) is borderline by the FDR threshold (ECB-only post-1999). At monthly cadence the per-step velocity coefficient of variation is $\sim$1--2\% across all indicators, consistent with entropic-GW thermal equipartition. Regime detection is therefore a quarterly-or-coarser signal, not a monthly one. 

\begin{figure*}[!tp]
\centering
\includegraphics[width=\textwidth]{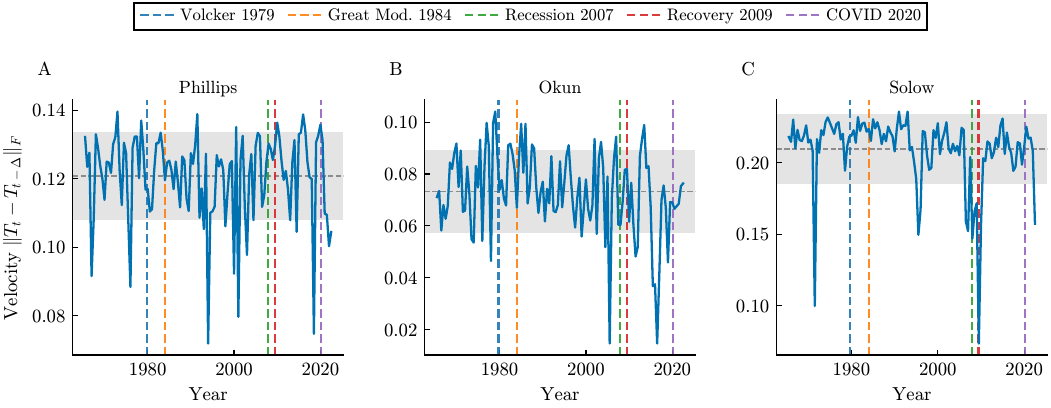}
\caption{\textbf{Transport-plan velocity is small and stable, with canonical regimes marked individually.} Frobenius velocity $v(t)=\lVert T_t-T_{t-\Delta}\rVert_F$ of the rolling-window transport plan to each law's full-sample reference attractor, on a 6-month grid (1965--2022), for Phillips (\textbf{A}), Okun (\textbf{B}), and Solow (\textbf{C}). The dashed line is the mean and the shaded band is the standard deviation. Colored vertical lines mark five canonical US regime dates (Volcker 1979, Great Moderation 1984, Great Recession 2007, recovery 2009, COVID 2020). The velocity stays within a tight band across six decades. Its first four moments per law are tabulated in Table~\ref{tab:velmom}. }
\label{fig:velmom}
\end{figure*}

We find that velocities are small and stable. The regime steps above are rare events. Most of the time the dynamical basis barely moves but it does move when there are materially large changes to the economy. Measuring $v(t)$ on rolling windows for each law (Fig.~\ref{fig:velmom}) confirms it. The velocity sits in a tight band with coefficient of variation $0.11$ (Phillips), $0.22$ (Okun), and $0.12$ (Solow), so the relational geometry is stable to order ten to twenty percent across six decades. The distributions are left-skewed (skewness $-1.4$, $-0.9$, $-2.9$) and strongly leptokurtic (excess kurtosis $2.6$, $1.7$, $10.3$). The basis moves at a near-constant baseline rate punctuated by occasional quiescent lulls rather than by large bursts, which is why a raw velocity threshold detects regimes poorly and the standardized permutation-null step above is the better signal. The first four moments of $v(t)$ (Table~\ref{tab:velmom}) are themselves compact econometric features of a series' dynamical stability. Parameters: $N_{\mathrm{ref}}=N_{\mathrm{win}}=20$ supports, entropic-GW $\varepsilon=0.02$, 6-month grid 1965--2022. 

\begin{table}[t]
\centering
\caption{\textbf{First four moments of the transport-plan velocity $v(t)$ per law.} Mean, coefficient of variation (CV), skewness, and excess kurtosis over the 1965--2022 grid (averaged over 8 entropic-GW initialization seeds). Every law is stable (low CV), left-skewed, and leptokurtic.}
\label{tab:velmom}
\begin{ruledtabular}
\begin{tabular}{lcccc}
Law & Mean & CV & Skewness & Excess kurtosis \\
\hline
Phillips & $0.121$ & $0.11$ & $-1.44$ & $2.58$ \\
Okun & $0.073$ & $0.22$ & $-0.88$ & $1.74$ \\
Solow & $0.210$ & $0.12$ & $-2.88$ & $10.32$ \\
\end{tabular}
\end{ruledtabular}
\end{table}

\subsubsection{Transport-plan velocity and Chow-\texorpdfstring{$F$}{F} detect complementary classes of regime change}
\label{subsubsec:velocity-chow}

The Chow test~\cite{chow1960tests} is the textbook regime detector: split a series at a candidate date, fit a linear model on each side, and report an $F$-statistic for whether the regression coefficients differ. By construction it hunts for breaks in the mean relationship and is blind to breaks that change only the variance or the higher-moment geometry. It also requires the break date to be named in advance, which the modern literature removed. Andrews' sup-Wald statistic tests for a break of unknown timing~\cite{andrews1993tests}, and the Bai--Perron procedure estimates multiple breaks and their dates jointly~\cite{bai1998estimating,bai2003computation}, reviewed in \cite{hansen2001new}. We scan every admissible split point precisely so that the comparison is against the whole candidate set rather than a single assumed date, but we note that the sharper benchmark for future work is a sup-Wald or Bai--Perron statistic rather than Chow alone. Our transport-plan velocity assumes no linear model at all, measuring how fast the coupling geometry itself is moving.

The comparison across five canonical US regime dates (Volcker 1979-10, Great Moderation 1984-01, Great Recession 2007-12, recovery 2009-06, COVID 2020-03) is shown in Table~\ref{tab:chowf-ranks}. Ranks run over the 116 candidate break dates in each law's monthly window, every admissible split point, so a low rank means the canonical date stands out against every alternative date, not merely against a null. Neither detector dominates, and the pattern of wins is the result. The Hellinger velocity is decisive exactly where the linear detector is structurally blind. The Great Moderation is the canonical variance break, a change in volatility with little movement in the mean relationship \cite{mcconnell2000output}, and on Okun it ranks $34$ of $116$ for velocity against $76$ for Chow-$F$. The Great Recession ranks $3$ against $15$ on Okun and $38$ against $69$ on Phillips, and the recovery ranks $6$ against $71$ on Phillips. The reverse holds on Volcker. The disinflation was a deliberate break in the mean relationship, precisely the object the Chow regression is built to find, and there Chow-$F$ outranks velocity in both laws. COVID splits, with velocity ahead on Phillips and Chow-$F$ ahead on Okun. A mean-based test and a geometry-based test see different classes of break, and the dates each one wins are the dates its mechanism predicts it should win. Transport velocity catches the variance breaks because Gromov--Wasserstein couplings are sensitive to second-moment deformations of the attractor geometry. This is the practical payoff of the construction. A detector that watches the geometry of the phase space of phase spaces, rather than the coefficients of a fitted model, sees a regime change as what it physically is, a deformation of the dynamical basis, and it complements rather than replicates the mean-break tests econometrics already has. 

\begin{table*}[ht!]
\centering
\caption{\textbf{Velocity and Chow-$F$ ranks of five canonical US regime dates.} Rank of each canonical date among the $116$ admissible break dates in each law's monthly window for the Hellinger transport-plan velocity against the Chow-$F$ statistic, lower is stronger, with bold marking the better detector per law and date. Velocity wins the variance-flavored breaks and Chow-$F$ wins the mean break of the Volcker disinflation. Solow is omitted because its window yields no valid rolling plans on this grid. }
\label{tab:chowf-ranks}
\begin{ruledtabular}
\begin{tabular}{lcccc}
 & \multicolumn{2}{c}{Phillips} & \multicolumn{2}{c}{Okun} \\
Regime date & Velocity & Chow-$F$ & Velocity & Chow-$F$ \\
\hline
Volcker 1979 & $112$ & $\mathbf{47}$ & $71$ & $\mathbf{68}$ \\
Great Moderation 1984 & $18$ & $\mathbf{11}$ & $\mathbf{34}$ & $76$ \\
Great Recession 2007 & $\mathbf{38}$ & $69$ & $\mathbf{3}$ & $15$ \\
Recovery 2009 & $\mathbf{6}$ & $71$ & $46$ & $\mathbf{7}$ \\
COVID 2020 & $\mathbf{29}$ & $70$ & $81$ & $\mathbf{32}$ \\
\end{tabular}
\end{ruledtabular}
\end{table*}

\section{Discussion}
\label{sec:discussion}

The phase space of phase spaces provides a universal geometric basis for data-driven time-series problems. The reliance on dynamical systems priors such as Takens' Theorem allows for the faithful reconstruction of individual bases for any given timeseries while the use of the transport plans between such timeseries allows us to understand each of these attractors in conversation with each other. Having validated the architecture on simple chaotic dynamical systems and extended it to the $\sim$281{,}000 macroeconomic time series of the universal corpus, we have constructed a compact geometrical basis for the macroeconomy. Validating the embeddings of key macroeconomic relationships like the Phillips curve, Okun's Law and the Solow model, we have also demonstrated the inadequacy of linear techniques to perform these reconstructions. Finally, by tracking the velocity of these transport plans, whose first four moments are themselves compact econometric features of dynamical stability, we demonstrate that they detect the variance-flavored regime changes the standard Chow-$F$ test is structurally blind to, because they register deformations of the geometric structure of the phase space of phase spaces that a mean-based test cannot see.

The contribution reads differently in each field it touches. For macroeconomics it supplies a basis rather than a model. The canonical laws are not separate specifications but low-dimensional neighborhoods of one connected object, and a regime change becomes a measurable deformation of that object rather than a break in a fitted coefficient of the kind sought by Markov-switching~\cite{hamilton1989newapproach} or break-testing~\cite{andrews1993tests,bai2003computation} specifications. The transmission mechanism itself surfaces as a direction on that object, with productivity and the interest rate the net sources of the coupling network and credit and output its deepest sinks, the credit channel recovered without a single behavioral assumption~\cite{bernanke1995inside}. The Phillips relation itself is promoted from a contested slope to a measured topology, a persistent loop whose hysteresis is a computable invariant of the joint attractor rather than a story about expectations, and the IAAFT surrogates certify that linearization destroys this geometry.

For dynamical systems it extends the delay-embedding guarantee~\cite{takens1981detecting,sauer1991embedology}, which is a statement about one series, to a population of them. Stark's extension to forced and noisy systems~\cite{stark1999delay,stark2003delay} makes each reconstruction admissible individually, but says nothing about how two of them relate. The transport plan to a shared barycenter is what makes independently reconstructed attractors comparable at all. For geometry the object of interest is the submanifold of the Birkhoff polytope that the plans occupy. Gromov--Wasserstein supplies a distance between metric measure spaces~\cite{memoli2011gromov} and an entropic solver makes it computable~\cite{peyre2016gromov,cuturi2013sinkhorn}, but neither fixes the metric on the resulting family of plans. Ours is Hellinger rather than Frobenius, chosen for the ${\sim}n_s/2$ resolution amplification of \S\ref{sec:gw} and validated on the corpus itself, where Hellinger coordinates keep $98\%$ of series distinct at a threshold that collapses raw-plan coordinates into near-duplicates. For physics the claim is the familiar one that a strongly coupled high-dimensional system is governed by few slow modes~\cite{wilson1971renormalization,coifman2006diffusion}, tested here on an economy rather than a fluid: $281{,}536$ observed series collapse onto roughly seven effective dimensions, and phase-randomized surrogates~\cite{schreiber1996improved} destroy a quarter to a half of that structure, so the compactness is dynamical rather than statistical.

The construction inherits Takens' assumptions. It needs series long enough to sample their attractor and approximately stationary after the usual econometric preprocessing, so very short or violently non-stationary records sit outside its reach, the same boundary that limits nonlinear forecasting from delay coordinates generally~\cite{casdagli1989nonlinear}. The transport plans resolve structure only down to the entropic-regularization floor set by $(n_s,\varepsilon)$, so two nearly identical economies, French and German GDP say, can blur together. Every coordinate is conditional on a frozen barycenter and a chosen corpus: change the corpus and the coordinates move even when the underlying geometry does not. The conceptual-barycenter network depends on a human-chosen indicator set and an edge threshold, and we have not shown that an unsupervised, purely dynamics-driven choice of barycenters would reproduce the same source and sink structure. While the basis is a strong regime detector, prediction is not demonstrated here and remains the main open direction. The natural benchmark is the diffusion-index forecast built on dynamic factors~\cite{stock2002forecasting}, run on these coordinates instead of on linear ones and compared under a test valid for nested models~\cite{clark2007approximately}.

The natural next steps follow the same logic at finer grain. Because the phase space of phase spaces can be resolved at any timescale, building it on region- or instrument-specific data using Eurostat NUTS panels, limit-order books and so on, should yield bases tuned to whatever dynamics dominate that setting. Widening the source pool should, in turn, let the PSoPS grow sparser rather than denser, since each new stream mostly reinforces directions that already exist. The boldest version of the idea is cross-domain. Many sources of driving dynamics may appear similar in spite of vastly different subject matter. That we may use Poisson statistics to describe both photon emission and bus arrivals at a given stop comes naturally to us. The same logic may be applied to families of dynamics. Navier--Stokes and Black--Scholes share a diffusive backbone, so weather and options pricing ought to meet in a common geometry. Turning the basis from a detector into a forecaster will mean tuning the construction to specific classes of prediction problems.

\emph{Code availability.} All figures and numerical results are reproduced by self-contained scripts in the companion repository, \url{https://github.com/maxtopel/phase-space-of-phase-spaces}. The four transport-plan binaries read by the corpus-level results are deposited on Zenodo, \href{https://doi.org/10.5281/zenodo.22602656}{doi:10.5281/zenodo.22602656}. Each \texttt{Source:} pointer in the text names the script in that repository that regenerates the corresponding result.

\begin{acknowledgments}
This work has been about 5 years in the making. It started by applying my work with Andrew Ferguson at the University of Chicago to exchange rate and commodities dynamics that brought me to Jean-Philippe Bouchaud and Michael Benzaquen's EconophysiX lab where I was able to pursue it full time. It was only over the course of the last 2 years that it reached its current iteration as a basis construction algorithm for arbitrary dynamics. It was my collaboration with Madhav Mani at Northwestern University on embryology that provided the final pieces of topological thinking to understand the proper architectural decisions required to make this project interpretable and my conversations and work with Omer Bensaadon at PSoPS that brought this from theory to practice. To all involved in this project at any level but especially to these key collaborators over the years, I am eternally grateful.

This work rests on openly published data, used with attribution under each provider's terms: Eurostat (\copyright~European Union), the European Central Bank, the World Bank, the Penn World Table~\cite{feenstra2015next}, the OECD, the Bank of England, the Bank for International Settlements, the International Monetary Fund, FRED of the Federal Reserve Bank of St.\ Louis, the Bank of Japan, the Bank of Canada, the Reserve Bank of Australia, the Swiss National Bank, and the public health, weather and hydrology archives of the WHO, NOAA and national river-gauge networks.
\end{acknowledgments}

\clearpage

\section{Corpus composition}\label{sec:corpus}

Table~\ref{tab:corpus-sources} enumerates the ingested corpus by provider, Table~\ref{tab:corpus-geography} its reporting-entity and frequency composition, and Table~\ref{tab:corpus-subsets} the gated subsets entering each analysis.

\begin{table}[!h]\centering
\caption{\textbf{Data sources.} The full ingested corpus is $281{,}536$ series across 17 providers. Source: \texttt{atlas/data/streams/economics/corpus\_manifest.json}.}
\label{tab:corpus-sources}
\begin{ruledtabular}\begin{tabular}{lr}
Provider & Series \\ \hline
\texttt{eurostat} & $219{,}410$ \\
\texttt{ecb} & $32{,}219$ \\
\texttt{worldbank} & $16{,}468$ \\
\texttt{pwt} & $6{,}849$ \\
\texttt{oecd} & $1{,}741$ \\
\texttt{who} & $1{,}335$ \\
\texttt{boe} & $1{,}183$ \\
\texttt{bis} & $965$ \\
\texttt{imf} & $600$ \\
\texttt{yahoo} & $499$ \\
\texttt{fred} & $106$ \\
\texttt{rba} & $45$ \\
\texttt{snb} & $34$ \\
\texttt{rivers} & $30$ \\
\texttt{noaa} & $22$ \\
\texttt{boc} & $19$ \\
\texttt{boj} & $11$ \\
\hline Total & $281{,}536$ \\
\end{tabular}\end{ruledtabular}\end{table}

\begin{table}[!h]\centering
\caption{\textbf{Reporting-entity and frequency composition of the ingested corpus.} The country field mixes ISO country codes with World Bank aggregate codes, statistical aggregates and UN M49 region codes. Net of the aggregate codes and of duplicates across the two ISO code systems, the corpus covers $232$ distinct national economies, $222$ of them within the economic sources.}
\label{tab:corpus-geography}
\begin{ruledtabular}\begin{tabular}{lrl}
Class & Distinct & Example \\ \hline
ISO-3 country codes & $232$ & \texttt{USA}, \texttt{JPN} \\
ISO-3 World Bank aggregates & $48$ & \texttt{WLD}, \texttt{HIC} \\
ISO-2 codes & $48$ & \texttt{FR}, \texttt{EA} \\
Statistical aggregates & $15$ & \texttt{EU}, \texttt{EA19}, \texttt{G7} \\
Regional groupings & $76$ & \texttt{EUROPE}, \texttt{LLDC} \\
UN M49 numeric codes & $40$ & \texttt{142}, \texttt{150} \\
\hline
Frequency: monthly & $171{,}682$ & \\
Frequency: quarterly & $70{,}610$ & \\
Frequency: annual & $38{,}329$ & \\
Frequency: daily & $915$ & \\
\end{tabular}\end{ruledtabular}\end{table}

\begin{table*}[!ht]\centering
\caption{\textbf{Corpus subsets used in this paper.} Each row is the population entering a specific analysis. No near-duplicate gate is applied at any stage: series whose transport plans coincide are distinct observations of the same dynamics and are retained. Source: \texttt{pipeline/corpus.py}, \texttt{pipeline/coordinates.py}.}
\label{tab:corpus-subsets}
\begin{ruledtabular}\begin{tabular}{llrl}
Subset & Gate & $N$ & Used for \\ \hline
Ingested corpus & none & $281{,}536$ & source enumeration \\
All-frequency & quality & $35{,}224$ & frequency pooling \\
High-resolution & $n\ge 200$ & $11{,}928$ & robustness check, $d_{\mathrm{eff}}=8.4\pm0.3$ \\
Distributed & $n\ge 60$ & $17{,}025$ & \textbf{primary corpus}, $d_{\mathrm{eff}}=7.3\pm0.2$ \\
Concept-capped & cap $12$\% & $7{,}864$ & composition invariance \\
Density-equalized & resampled & $10{,}133$ & composition invariance \\
FPS landmarks & --- & $800$ & Nystr\"om eigenbasis \\
Out-of-domain controls & --- & $156$ & held out, domain-scope test \\
\end{tabular}\end{ruledtabular}\end{table*}

\clearpage

\bibliography{bib}

\end{document}


\title{Supplementary Information for\\ Phase spaces of phase spaces: reconstruction of the geometry of the macroeconomy}

\author{Maximilian Topel}
 \affiliation{Department of Applied Mathematics, Northwestern University, 2145 Sheridan Road, Evanston, Illinois 60208, USA}

\date{\today}
\maketitle

\section{Why the transport plan and not its centroid}
A natural first instinct is to summarize each series by the centroid of its transported mass on the barycenter. The following remark is why the paper works with the full plan instead.

\begin{remark}[Centroid collapse]
\label{rem:centroid-collapse}
The na\"ive projection $\bar{c}_i = \frac{1}{n_s}\sum_j (T_i Z_b)_j$ (where $Z_b$ are MDS coordinates of the barycenter) is identically zero for all series.
This is a mathematical identity rather than a numerical artifact, since $T_i$ has uniform column marginals $q_j = 1/n_s$ and MDS coordinates are centered ($\sum_j Z_b(j,:) = 0$),
\[
  \bar{c}_i = \sum_j q_j Z_b(j,:) = \frac{1}{n_s}\sum_j Z_b(j,:) = 0.
\]
The centroid depends only on the column marginal $q$ and the barycenter mean, not on the transport plan~$T_i$.
The transport plans are informative (empirically, $\|T_i - \frac{1}{n_s^2}\mathbf{1}\|_F \sim 0.05$--$0.09$ at $n_s = 20$, an early benchmark configuration), but the centroid summary discards all structure.
\end{remark}

\section{Stationarity and Preprocessing}
\label{sec:stationarity}

\subsection{Three-Stage Classification}

The first stage detects structural breaks.
We apply the PELT algorithm \citep{killick2012optimal} with $L_2$ cost and BIC penalty. Like the Bai--Perron dynamic-programming procedure for multiple structural breaks \citep{bai2003computation}, PELT returns an exact global minimiser of the penalised cost, but prunes the candidate set to achieve linear rather than quadratic cost in the series length.
If a break is detected, the series is split at the breakpoint and each segment is classified independently.
The most recent segment determines the primary classification.

The second stage classifies integration order.
We use the ADF/KPSS joint protocol \citep{dickey1979distribution,kwiatkowski1992testing}, which yields a four-cell classification matrix:
\begin{center}\small
\begin{tabular}{c|cc}
  & KPSS does not reject & KPSS rejects \\
\hline
ADF rejects & $I(0)$ (stationary) & Inconclusive \\
ADF does not reject & Inconclusive & $I(1)$ (unit root) \\
\end{tabular}
\end{center}

For the ``inconclusive'' cells, we estimate the fractional integration parameter $d$ via the GPH estimator \citep{geweke1983estimation}.
Following \citet{baillie1996long}, a series with $\hat{d} < 0.5$ is left in levels so its long memory is preserved, and a series with $\hat{d} \ge 0.5$ is treated as $I(1)$ and difference.

For monthly and quarterly series, we apply the HEGY test \citep{hylleberg1990seasonal} for seasonal unit roots.
If a seasonal unit root is detected at a non-zero frequency but ADF finds $I(0)$ at zero frequency, we apply seasonal differencing $(1 - L^s)$ before the standard routing.

The third stage routes each series to its transform.
A deterministic table maps (economic type $\times$ integration order) to a specific transform:

\begin{center}\small
\resizebox{\columnwidth}{!}{%
\begin{tabular}{lcccc}
\toprule
Type & $I(0)$ / Frac$(d<0.5)$ & $I(1)$ / Frac$(d \ge 0.5)$ & $I(2)$ \\
\midrule
Level (GDP, CPI) & $\log$, leave & $\log$, $\Delta$ & $\log$, $\Delta^2$ \\
Rate (interest) & leave & $\Delta$ & $\Delta^2$ \\
Index (PMI) & leave & $\Delta$ & $\Delta$ \\
Physical (temp) & deseason, leave & deseason, $\Delta$ & deseason, $\Delta$ \\
\bottomrule
\end{tabular}}
\end{center}

\subsection{What Differencing Does to Attractors}

Differencing changes the data-generating process. If $\{x_t\}$ has attractor $\calA$, the differenced series $\{y_t = x_t - x_{t-1}\}$ has a different attractor $\calA'$.
For $I(1)$ processes, $\calA$ is non-compact (the random walk fills $\R$), so Takens' theorem \citep{takens1981detecting} does not apply.
Differencing restores compactness and stationarity, enabling meaningful embedding, at the cost of losing the level information.

\subsection{Scalability and Parameter Selection}

For $n > 1000$ series, we use Nystr\"om-DMAP. We select $K = 800$ landmarks via FPS in Hellinger space, compute the dense $K \times K$ diffusion map, and extend to all $n$ series via the Nystr\"om formula \citep{williams2001nystrom}, the standard out-of-sample extension for spectral embeddings of this type \citep{bengio2004outofsample}.
This reduces complexity from $O(n^2)$ to $O(nK)$ and amortizes the dense spectral embedding only. Per-series GW is still solved for every series.
The memory footprint is the $n \times n_s^2$ transport-plan array at float32, which for 281K series at $n_s = 40$ comes to $281{,}000 \times 1{,}600 \times 4$ bytes, about 1.8\,GB.
We verify the extension reproduces the dense embedding on both clean and real data. With FPS landmarks in the graded-similarity regime, the Nystr\"om coordinates recover the full dense Hellinger-DMAP's leading modes with per-mode correlation $\rho_1 = 0.97$, $\rho_2 = 0.93$ on the macroeconomic monetary domain (1200 series), and $\rho = 0.88$ across all modes on the chaotic-systems corpus (landmark self-consistency $\rho = 1.0$).
Two conditions are required for this to work. The landmarks must come from farthest point sampling, since random landmarks degrade the reconstruction sharply. And the regularization must sit in a regime of graded plan similarity, because in the near-orthogonal regime (cost/$\varepsilon$ too large) the Hellinger-DMAP has no smooth manifold to extend and the correlation collapses. On the monetary domain $\rho_1$ falls from $0.97$ to $0.13$ as the mean pairwise Hellinger spread between plans rises from $0.55$ to $0.97$. This is the same graded-similarity requirement identified below for the GW regularization itself (\Cref{sec:gw-entropic-exact}).
Source: \texttt{validation/nystrom\_equivalence.py}, \texttt{validation/nystrom\_monetary.py}.

We also benchmarked the landmark approximation against exhaustive GW directly. On seven macroeconomic test corpora (Phillips, Okun, Solow, PPP, monetary transmission, global macro and Fama--French, with $N = 250$--$500$ series each) we reconstruct the full $N \times N$ exhaustive per-series GW distance matrix from $K$ landmarks (FPS/random sampling $\times$ triangle/Euclidean/Nystr\"om reconstruction) and compare to the exhaustive ground truth by Frobenius relative error, pairwise-distance Spearman~$\rho$, and leading-mode (mode-1) cosine similarity.
The dominant PSoPS (phase space of phase spaces) mode, which sets the coordinates, transfers well for FPS--Nystr\"om at $K = 50$, with mode-1 cosine $0.99$ (Okun), $0.995$ (global macro, monetary transmission), $0.97$ (PPP).
The reconstruction is, however, method- and corpus-dependent.
Random-landmark variants and the Solow and Phillips corpora can degrade sharply (mode-1 cosine $0.04$--$0.5$ for the wrong method/corpus combination), and the full pairwise-distance Spearman~$\rho$ reaches only $0.66$--$0.83$ even for the best method.
The landmark approximation therefore faithfully transfers the dominant geometry but loses finer pairwise structure. We use FPS landmarks and validate the resulting landscape directly via the split-half convergence battery (Procrustes disparity $0.09$, \Cref{sec:convergence}) rather than relying on distance-matrix reconstruction alone.
Source: \texttt{data/atlas\_anchors/\allowbreak results/\allowbreak aoa\_39\_landmark\_benchmark.json}.

The support size $n_s$ controls a fundamental trade-off. Smaller $n_s$ gives more structured transport plans (larger deviation from uniform, stronger GW signal) but coarser attractor representation (fewer skeleton points).
At $n_s = 200$, transport plans degenerate to near-uniform regardless of epsilon, while at $n_s = 10$, deviations reach $\sim$9\% of the Frobenius norm.
We sweep $n_s \in \{10, 15, 20, 30, 33, 35, 37, 40, 43, 45, 47, 50, 60, 75\}$ and select based on DMAP spectral gap $\lambda_1/\lambda_2$.
On 100 real economic series, a phase transition occurs at $n_s \approx 35$, where the spectral gap jumps from $1.37$ (at $n_s = 33$) to $2.39$ (at $n_s = 35$).
The plateau $n_s \in [35, 60]$ maintains gap $> 1.86$ with effective dimension $d_{\mathrm{eff}} \approx 6$--$10$, a bracket consistent with the final headline $d_{\mathrm{eff}}=7.3\pm0.2$ of the main text.
Bootstrap stability analysis (10 replicates $\times$ 80-series random subsets) reveals that $n_s = 40$ has the lowest coefficient of variation (CoV $= 0.13$, min gap $= 1.66$), compared to $n_s = 35$ (CoV $= 0.37$, min gap $= 1.02$).
We set $n_s = 40$, the most stable spectral gap, giving $40^2 = 1{,}600$ transport plan entries per series. The sweep is recorded in the tuning provenance shipped with the pipeline configuration (\texttt{vendor\_atlas/config.py}).

The regularization $\varepsilon$ was tuned alongside it. In the legacy alignment layer the entropic regularization $\varepsilon$ was swept over $\{0.01, 0.05, 0.1, 0.5\}$. The production barycenter is fitted without entropic regularization, and entropy enters only in the per-series plans onto the frozen reference.
For the per-series GW transport, we use $\varepsilon = 0.008$ on unit-scaled costs (small enough to preserve structure, large enough to keep $T_i$ in the interior of $\mathcal{B}_{n_s}$, which is required for the Hellinger metric).

\subsection{Entropic vs.\ exact Gromov--Wasserstein}
\label{sec:gw-entropic-exact}

Whether the transport plans carry usable structure hinges on how the GW problem is regularized.
Exact (unregularized) GW returns sharp, near-permutation plans. On the controlled chaotic-systems corpus, projecting every series to a common reference attractor and measuring the mean pairwise Hellinger distance between the resulting plans gives $0.976$ (range $0.82$--$1.00$), so plans for distinct attractors are almost orthogonal.
Entropic GW instead returns a smooth, strictly-interior plan (required for the Hellinger metric on $\mathcal{B}_{n_s}$), but at fixed $\varepsilon$ the entropy term competes with the cost term, so the operative control is the ratio of cost magnitude to $\varepsilon$.
Holding $\varepsilon = 0.05$ and scaling the unit-normalized cost matrices by a factor $s$, the mean pairwise Hellinger spread rises from $0.001$ at $s = 1$ (every plan collapses onto the uniform coupling) through $0.034$ at $s = 3$ to $0.859$ at $s = 6$, a sharp transition out of the entropic limit into the structured regime.
We therefore operate at a cost/$\varepsilon$ ratio inside the structured regime, equivalently $\varepsilon \sim 0.008$ on unit-scaled costs, small enough to preserve plan structure, large enough to keep $T_i$ interior.
This is the same degeneracy that appears at large support size ($n_s = 200$ collapses to near-uniform regardless of $\varepsilon$).
Source: \texttt{validation/nystrom\_equivalence.py}.

\section{Convergence Theory}
\label{sec:convergence}

\subsection{GW Barycenter Convergence}

The GW barycenter is the solution of a non-convex optimization problem.
Global convergence is not guaranteed. Practical convergence depends on initialization and $\varepsilon$.
The production barycenter is fitted without entropic regularization from a random initialization, and its stability is checked directly by the split-half convergence battery (Procrustes disparity $0.09$, \texttt{validation/split\_half.py}) rather than by multi-initialization comparison.

\subsection{Sample Complexity: How Many Series?}

The central empirical question is how many series $N$ are needed before the landscape stabilizes.
We measure stability via three complementary metrics:

\begin{enumerate}
  \item \textbf{Procrustes distance.} After optimal rotation and scaling \citep{gower1975procrustes}, the residual between the $N$-series landscape and the full landscape. Threshold: $< 0.05$.
  \item \textbf{Eigenvalue ratio distance.} $\sqrt{\frac{1}{r}\sum_{k=1}^r (\lambda_k^{(N)}/\lambda_1^{(N)} - \lambda_k^{(\text{full})}/\lambda_1^{(\text{full})})^2}$ where $r=10$. Threshold: $< 0.02$.
  \item \textbf{Probe rank correlation.} Spearman $\rho$ of pairwise distances among fixed probe points, with threshold $> 0.98$.
\end{enumerate}

Convergence is judged by a running-mean stability criterion rather than by Geweke, CUSUM, or ADF tests on the metric sequences, which test time-series stationarity rather than manifold convergence. A metric is converged when the absolute change in its running mean over the last 10 replications falls below $5\%$ of the overall standard deviation, and formal convergence is declared when the eigenvalue ratio changes by less than $2\%$ for 3 consecutive batch sizes across 30 replications.

\subsection{Bootstrap Validation}

We use 500 bootstrap replicates with 60\% subsamples rather than 80\%, since at 80\% the subsamples are highly correlated and inflate coverage estimates.
At 100 replicates, coverage estimates have $\pm 4.4$pp uncertainty. 500 replicates reduce this to $\pm 2.0$pp.

\subsection{Intrinsic Dimensionality Estimation}

Alongside the Coifman $\varepsilon$-scaling estimator of record, we track four secondary estimates of the effective dimensionality $d_{\mathrm{eff}}$ of each landscape:

\begin{enumerate}
  \item \textbf{TWO-NN} \citep{facco2017estimating}, the most robust of the four, needs no scaling region. It is based on the ratio $\mu = r_2/r_1$ of second to first NN distance, giving $d = n / \sum \log \mu_i$.
  \item \textbf{Correlation dimension $D_2$}, the classical Grassberger--Procaccia estimate with a convergence gate.
  \item \textbf{Levina--Bickel MLE} \citep{levina2004maximum}, the local maximum-likelihood dimension averaged over all points.
  \item \textbf{Participation ratio}, $\mathrm{PR} = (\sum \lambda_i)^2 / \sum \lambda_i^2$, from PCA eigenvalues.
\end{enumerate}

Nearest-neighbor maximum-likelihood estimators, TWO-NN~\citep{facco2017estimating} and Levina--Bickel~\citep{levina2004maximum}, infer dimension from short-range distance statistics and are sensitive both to sampling density and to the number of retained components. Our scaling sweep on the universal corpus shows the failure directly. Growing the reference corpus from $10$ to $8000$ series moves TWO-NN only from $10.80$ to $10.93$ (sweep minimum $10.36$ at $N=25$), and holds Levina--Bickel between $17.3$ and $20.5$ against a $20$-component basis. An estimator that ignores an eight-hundred-fold increase in data is reading the ambient embedding rather than the manifold. The spectral variance counts saturate the same way, sitting at the retained-component ceiling. The main text therefore quotes the bandwidth-free Coifman $\varepsilon$-scaling dimension.

The Coifman estimate itself should be read as a floor. The exponent is exact only for a densely sampled manifold. At finite $n$ the scaling window is bounded below by the self-pair regime and above by kernel saturation, and that window narrows as the dimension grows, because populating a $d$-ball takes exponentially many points. The estimator therefore reads low, increasingly so with $d$. Calibrating it against $d$-spheres of known dimension embedded in the same $1600$-dimensional ambient space and evaluated at the same sample sizes, a true $5$-manifold reads $4.60$, a $7$-manifold $5.97$, a $9$-manifold $7.01$ and an $11$-manifold $7.81$. The corpus reads $7.19$ at that sample size, which is what a nine-dimensional object reads, and inverting the calibration curve implies a true intrinsic dimension near $9.5$. The main text quotes the measured $7.3$ rather than the corrected value, because the calibration uses uniformly sampled spheres and the corpus is neither uniform nor spherical, and non-uniform sampling widens the deficit rather than narrowing it. Either way, the manifold cannot be lower-dimensional than about nine. Source: \texttt{validation/convergence\_floor.py}, which writes \texttt{artifacts/\allowbreak convergence\_study.json}.

\section{Macroeconomic Validation}
\label{sec:macro-validation}

Each law is tested geometrically. The main text reports the three canonical laws (Phillips, Okun, Solow) in depth.

\subsection{Geometric Tests}

Three geometric tests apply. For directed chains such as monetary transmission, running from the policy rate through short and long rates, credit spreads, lending and output to inflation, we test sequential proximity, $d(s_i, s_{i+1}) < d(s_i, s_{i+2})$ where $s_i$ is the centroid of the step-$i$ series. For co-determined pairs such as inflation and unemployment we test neighborhood overlap, asking whether members of one group are over-represented among the $k$-nearest neighbors of the other under the hypergeometric distribution. And for production functions such as the Solow $Y$--$K$--$L$ triangle we test whether the geometric arrangement of the factor centroids is consistent with production-function predictions, for example capital sitting closer to output than labor does in capital-intensive economies.

Across these tests, $p$-values within each time period are corrected via Benjamini--Hochberg \citep{benjamini1995controlling} at FDR level $q = 0.05$, and we report both raw $p$-values and BH-adjusted $q$-values.

\subsection{Okun's Law (Primary Specification)}

We test the gap version of the law, the HP-filtered output gap ($\lambda = 1600$ for quarterly data \citep{hodrick1997postwar} and $\lambda = 129{,}600$ for monthly \citep{ravn2002adjusting}) against the unemployment gap.
The difference version ($\Delta Y$ vs.\ $\Delta U$) serves as a robustness check.

The landscape test asks that output-gap series and unemployment-gap series be proximate in the GW landscape (neighborhood overlap) and their landscape distance should correlate with the Okun coefficient $\beta$ across countries.

\section*{Appendix: Validation on Known Dynamical Systems}
\label{sec:validation-dysts}

We validate the DMAP and descriptor pipeline on three systems with known properties:

\begin{center}
\resizebox{\columnwidth}{!}{%
\begin{tabular}{lcccl}
\toprule
System & True $d$ & True $D_2$ & True $\lambda_{\max}$ & Topological features \\
\midrule
Lorenz & 3 & $\approx 2.05$ & $\approx 0.91$ & Strange attractor, $H_0$ \\
R\"ossler & 3 & $\approx 1.99$ & $\approx 0.07$ & Band attractor \\
2-torus & 2 & 2.0 & 0 & Two independent $H_1$ 1-cycles \\
\bottomrule
\end{tabular}}
\end{center}

The measured recovery on these systems is reported in the main text. From a single scalar observable, the delay reconstruction returns $D_2$ within estimator uncertainty of the same estimator run on the true state for both Lorenz and R\"ossler, and persistent homology detects the $\beta_1=2$ torus topology.

\bibliography{bib}